\documentclass[preprint,aps,tightenlines,eqsecnum,floatfix,showpacs]{revtex4}
\usepackage{dcolumn}
\usepackage{bm}
\usepackage{color}
\usepackage{graphicx}
\usepackage[active]{srcltx}
\usepackage{pstricks}
\begin{document}

\title{Electromagnetic Currents and Magnetic Moments in $\chi$EFT}
\author{S.\ Pastore$^{\,{\rm a}}$, L.\ Girlanda$^{\,{\rm b,c}}$, R.\ Schiavilla$^{\,{\rm a,d}}$,
M.\ Viviani$^{\,{\rm c}}$, and R.B.\ Wiringa$^{\,{\rm e}}$}
\affiliation{
$^{\rm a}$\mbox{Department of Physics, Old Dominion University, Norfolk, VA 23529, USA}
$^{\,{\rm b}}$\mbox{Department of Physics, University of Pisa, 56127 Pisa, Italy}\\
$^{\,{\rm c}}$\mbox{INFN-Pisa, 56127 Pisa, Italy}\\
$^{\rm d}$\mbox{Jefferson Lab, Newport News, VA 23606}\\
$^{\,{\rm e}}$\mbox{Physics Division, Argonne National Laboratory, Argonne, IL 60439}\\
}

\date{\today}

\begin{abstract}
A two-nucleon potential and consistent electromagnetic currents are derived
in chiral effective field theory ($\chi$EFT) at, respectively, $Q^{\, 2}$
(or N$^2$LO) and $e\, Q$ (or N$^3$LO), where $Q$ generically
denotes the low-momentum scale and $e$
is the electric charge.  Dimensional regularization is used to
renormalize the pion-loop corrections.  A simple expression is
derived for the magnetic dipole ($M1$) operator associated
with pion loops, consisting of two terms, one of which is determined, uniquely, by
the isospin-dependent part of the two-pion-exchange potential.  This
decomposition is also carried out for the $M1$ operator arising from
contact currents, in which the unique term is determined by the
contact potential.  Finally, the low-energy
constants (LEC's) entering the N$^2$LO potential are
fixed by fits to the $np$ S- and P-wave
phase shifts up to 100 MeV lab energies.
\end{abstract}


\maketitle
\section{Introduction, conclusions, and outlook}
\label{sec:intro}
A quantitative understanding of low-energy nuclear physics in terms
of {\it ab initio} calculations in quantum chromodynamics (QCD) is
still lacking because of the non-perturbative character of the theory
in this regime.  However, the chiral symmetry exhibited by QCD
severely restricts the form of the interactions of pions
among themselves and with other particles~\cite{Weinberg95}.  In particular,
the pion couples to the baryons, such as nucleons or $\Delta$-isobars,
by powers of its momentum $Q$, and the Lagrangian describing
these interactions can be expanded in powers of $Q/\Lambda_\chi$, where
$\Lambda_\chi \sim 1$ GeV specifies the chiral-symmetry breaking
scale.  As a consequence, classes of Lagrangians emerge, each characterized
by a given power of $Q/\Lambda_\chi$ and each involving a certain number of
unknown coefficients, so called low-energy constants (LEC's), which are then
determined by fits to experimental data (see, for example, the
review papers~\cite{Bernard95} and~\cite{Epelbaum08}, and
references therein).

This approach, known as chiral effective field theory ($\chi$EFT),
has been used to study two- and many-nucleon interactions~\cite{Epelbaum08}
and the interaction of electroweak probes with nuclei~\cite{Park93,Park96}.
Its validity, though, is restricted to processes occurring at low energies.
In this sense, it has a more limited range of applicability than meson-exchange
or more phenomenological models of these interactions, which in fact
quantitatively and successfully account for a wide variety of nuclear
properties and reactions up to energies, in some cases, well beyond
the pion production threshold (for a review, see Ref.~\cite{Carlson98}).
However, it can be justifiably argued that
$\chi$EFT puts nuclear physics on a more fundamental basis
by providing, on the one hand, a direct connection between
QCD and its symmetries, in particular chiral symmetry, and the
strong and electroweak interactions in nuclei, and, on the
other hand, a practical calculational scheme susceptible,
in principle, of systematic improvement.

Recently we derived the nuclear electromagnetic current in a
$\chi$EFT with explicit pion, nucleon, and
$\Delta$-isobar degrees of freedom~\cite{Pastore08}.  Formal
expressions up to one loop---N$^3$LO or $e\, Q$ in the power
counting scheme, $Q$ generically indicating the low momentum
scale, and $e$ being the electric charge---were obtained in time ordered
perturbation theory (TOPT) by employing non-relativistic Hamiltonians implied
by the chiral Lagrangian formulation of Refs.~\cite{Weinberg90,vanKolck94,Epelbaum98}.
An important aspect of the derivations in Ref.~\cite{Pastore08} is in
the treatment of the reducible diagrams: recoil corrections, which arise
from expanding the nucleon energy denominators in these diagrams,
were found to partially cancel out the contributions from the irreducible
diagrams.  When applied to the nucleon-nucleon ($NN$) case, this approach
removes explicit energy dependencies, and in fact leads, at least up
to one loop (N$^2$LO or $Q^{\,2}$), to the same potential constructed by
Epelbaum {\it et al.}~by the method of the unitary transformation~\cite{Epelbaum98}.
It also generates N$^3$LO currents, which satisfy current conservation with
this potential.

One-loop $\chi$EFT currents have also been derived, with nucleons and
pions only, within the heavy-baryon (HB) formalism by Park {\it et al.}
in Ref.~\cite{Park96}, and have been used in calculations of the
$n$-$p$~\cite{Park96,Song07} and $n$-$d$~\cite{Song08} capture
cross sections, spin observables in $\vec{n}$-$\vec{p}$ capture~\cite{Park00},
and magnetic moments of the deuteron and trinucleons~\cite{Song07}, by
evaluating the relevant transition matrix elements between wave
functions obtained from realistic potentials, {\it i.e.}~in the
hybrid approach.  Later in the present work we shall show that
there are differences between the currents obtained in the HB
and TOPT formalisms, some of which have to do with the treatment
of reducible diagrams mentioned above (see Sec.~\ref{sec:mum}).
We should note that electromagnetic currents in the isoscalar
sector were also discussed in Refs.~\cite{Walzl01,Phillips03}, and
used in calculations of the deuteron static properties and elastic form
factors.

The N$^2$LO ($e\, Q^{\,0}$) currents, namely without loop corrections,
were used in Ref.~\cite{Pastore08} to carry out hybrid calculations
of the magnetic moments of $A$=2 and 3 nuclei, and thermal neutron
radiative captures on protons and deuterons.  To have an estimate of
the model dependence due to short-range phenomena, the variation
of the predictions was studied as a function of the cutoff parameter $\Lambda$,
needed to regularize the two-body operators entering the matrix elements,
as well as a function of the input potentials.  These N$^2$LO calculations did
not provide a satisfactory description of the experimental data, particularly
for the suppressed process $^2$H($n,\gamma$)$^3$H, which exhibited a
pronounced sensitivity to variations in $\Lambda$.  This clearly pointed
to the need of including loop corrections.

This work represents the next stage in the program initiated
in Ref.~\cite{Pastore08}.  It constructs, consistently within
the $\chi$EFT framework, a $NN$ potential and one- and two-body
currents up to N$^3$LO, with the ultimate aim of studying electromagnetic
properties and radiative captures in few-nucleon systems at this order.
More specifically, it fulfills two objectives.  The first is the construction, in dimensional
regularization, of a $NN$ potential at one loop (Sec.~\ref{sec:potential}).
The nine LEC's---$C_S$ and $C_T$ at $Q^{\,0}$, and $C_1,\dots,C_7$ at
$Q^{\,2}$ in the notation of Ref.~\cite{Epelbaum98}---which enter
the potential at this order are determined by fitting the $np$ S- and P-wave
phase shifts up to 100 MeV lab energies, obtained in the
recent partial-wave analysis of Gross and Stadler~\cite{Gross08} (Sec.~\ref{sec:fitp}).
Differences between the present version of the potential and
that obtained by Epelbaum and collaborators~\cite{Epelbaum00}
are not substantive, since they relate to the use of a different
form for the regulator in the solution of the Lippmann-Schwinger equation
and the adoption, in their case, of the older Nijmegen phase-shift analysis~\cite{Stoks93}
for the determination of the LEC's.

The second objective is to carry out the renormalization (in dimensional regularization)
of the tree-level and one-loop two-body currents, and to derive the complete set of contact
currents at N$^3$LO (Sec.~\ref{sec:currents}).  Those implied by minimal substitution
in the contact interaction Hamiltonians with two gradients of the nucleon fields were
in fact obtained in Ref.~\cite{Pastore08}.  However, in that work non-minimal couplings
were not considered: we remedy that omission here.
Lastly, in the present study we also
derive (renormalized) expressions for the magnetic dipole ($M1$)
operator at N$^3$LO
(Secs.~\ref{sec:mum} and~\ref{sec:muct}).  We find it convenient to separate, in the contributions from loop
corrections, a term dependent on the center-of-mass position of the two nucleons~\cite{Dalitz54,Sachs48},
which is uniquely determined via current conservation by the isospin-dependent part
of the two-pion-exchange chiral potential, and a translationally-invariant term.
The latter has a different isospin structure than that of Ref.~\cite{Park96}
for the reason mentioned earlier.

This decomposition is carried out also for the $M1$ operator
generated by the N$^3$LO contact currents.  The center-of-mass
dependent term is related to the contact potential, specifically
the part of it which is momentum-dependent and therefore
does not commute with the charge operator.  However, the translationally
invariant contact $M1$ operator depends on two LEC's.  There are
also N$^3$LO (translationally invariant) $M1$ corrections
at tree level, involving one-pion exchange, which depend on
three additional LEC's.  These five LEC's could be fixed either
by reproducing a combination of nucleon and nuclear
data---for example, pion-photoproduction data on a single nucleon
along with the deuteron magnetic moment and cross section for
$np$ radiative capture at thermal energies---or by relying exclusively
on nuclear data---by fitting, in addition to the observables
mentioned earlier, also the trinucleon magnetic
moments and radii.  In this respect, we note that there appear to be no
three-body currents entering at N$^3$LO (namely, $e\, Q^{-2}$ in
$A$=3 systems)~\cite{Pastore09}.

The stage is now set for carrying out a consistent $\chi$EFT calculation of electromagnetic
properties and reactions in $A$=2--4 nuclei.  The thermal neutron $n$-$d$ and $n$-$^3$He
and keV $p$-$d$ captures are especially interesting, since the $M1$ transitions
connecting the continuum states to the hydrogen- and helium-isotope bound states are inhibited
at the one-body (LO) level.  As a result, the cross sections for these processes are significantly
enhanced by many-body components in the electromagnetic current operator~\cite{Marcucci05,Schiavilla92}.
Work along these lines is in progress.  However, it remains to be seen whether the N$^3$LO operators
derived in this study will reduce the sensitivity to short-range physics found in the N$^2$LO
hybrid calculations (for the $n$-$d$ case) referred to earlier, and bring theory into satisfactory
agreement with experiment.

\section{ {\it NN} potential at N$^2$LO}
\label{sec:potential}
This section deals with the construction of the {\it NN} potential
in $\chi$EFT up to order $Q^2$, or N$^2$LO.  It is derived by
retaining only pions and nucleons as degrees of freedom---the
inclusion of explicit $\Delta$-isobar degrees of freedom is deferred
to a later work~\cite{Pastore09}.  The formalism as well as the
techniques we adopt have already been described in Ref.~\cite{Pastore08},
and we will not reformulate them here.

\begin{figure}[bthp]
\includegraphics[width=5in]{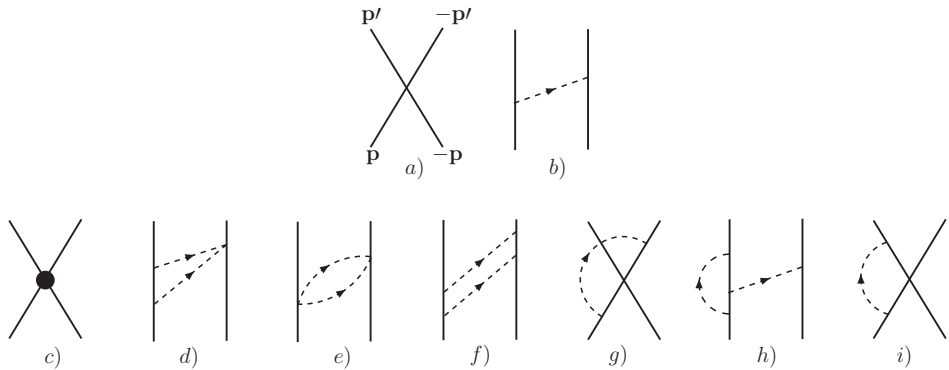}
\caption{Diagrams illustrating contributions to the $NN$ potential
entering at LO ($Q^{\,0}$), panels a) and b), and N$^2$LO ($Q^{\,2}$), panels c)-i).
Nucleons and pions are denoted by solid and dashed lines,
respectively.  The filled circle in panel c) represents the vertex from
contact Hamiltonians containing two gradients of the nucleons' fields.
Only one among the possible time orderings is shown for
each contribution with more than one vertex.}
\label{fig:fig1}
\end{figure}

In Fig.~\ref{fig:fig1} we show the diagrams illustrating the contributions
occurring up to N$^2$LO.  At LO ($Q^0$) there is a contact interaction,
panel a), along with the one-pion-exchange (OPE) contribution, panel b).
At N$^2$LO we distinguish among three different categories,
which are: i) contact interactions involving two gradients acting on the nucleons' 
fields, panel c); ii) two-pion-exchange loop contributions,
panels d)-f); and iii) loop corrections to the LO contact 
interaction, panels g) and i), and to the OPE contribution, panel h).
Note that in the figure we display only one among the possible time orderings.

The time ordered diagrams of panels a)-e) are irreducible, while those
of panels f)-g) have both reducible and irreducible character.
In order to avoid double counting of the reducible contributions
due to insertion of the LO potential into the Lippmann-Schwinger equation, the
{\it NN} potential is defined as the sum of the irreducible diagrams only.

The evaluation of the {\it NN} potential is carried out in the static limit.
Corrections to this approximation arise from kinetic energies of nucleons,
and are referred to as recoil corrections.  The latter are not
accounted for in the Lippmann-Schwinger equation in which only
the static potential is iterated.  Hence they have been included
below along with the irreducible contributions.  The resulting
potential is in agreement with that obtained by 
Epelbaum {\it et al.} in Ref.~\cite{Epelbaum98}.
Special treatment is reserved for the diagrams of panels h) and i),
which are discussed later in this section.
\subsection{Formal expressions}
\label{sec:pot_formulae}

In what follows we use the notation introduced in Ref.~\cite{Pastore08}.  In particular, the 
potential is obtained in the center-of-mass frame
where the nucleons' initial and final relative momenta
are ${\bf p}$ and ${\bf p}^{\prime}$, respectively.
We also define ${\bf k} = {\bf p}^{\prime} -{\bf p}$,
${\bf K}=({\bf p}^{\prime}+{\bf p})/2$,
$\omega_k= \sqrt{k^2+m_{\pi}^2}$, and
\begin{equation}
 \int_{{\bf p}} \equiv \int\frac{{\rm d}{\bf p}}{(2\pi)^3} \ .
\end{equation}
In the remainder of this section we will refer to the panels in Fig.~\ref{fig:fig1}.

The diagram illustrated by panel a) gives rise to the LO order contact potential
$v^{\rm CT0}$, which is expressed in terms of two LEC's  $C_S$ and $C_T$ as 
\begin{equation}
v^{\rm CT0}=C_S+C_T\,{\bm \sigma}_1\cdot{\bm \sigma}_2 \  ,
\label{eq:ct0}
\end{equation}
while that of  panel b) leads to the standard OPE potential,
\begin{equation}
\label{eq:ope}
v^{\pi}({\bf k})=-\frac{g_A^2}{F_{\pi}^2}\,{\bm \tau}_1\cdot{\bm \tau}_2\,
\frac{{\bm \sigma}_1\cdot{\bf k}\,{\bm\sigma}_2\cdot{\bf k}}{\omega_k^2}\  .
\end{equation}

Next we consider the contributions arising from panel c).
There is a number of contact Hamiltonians involving two 
gradients acting on the nucleons' fields compatible with the required 
symmetries of the underlying theory.  In fact, the list of fourteen
given in Ref.~\cite{Epelbaum98} and reported in Appendix~D
of Ref.~\cite{Pastore08} is redundant, since relations exist among the
terms proportional to $C^\prime_4$, $C^\prime_5$, and $C^\prime_6$, and
those proportional to $C^\prime_7$, $C_8^\prime$, $C^\prime_{10}$, $C^\prime_{11}$ (see Appendix~\ref{app:nmcounterterms}).
We will not enforce these in the following, since, in any case, the
contact Hamiltonians (all twelve of them) lead
(in the center-of-mass frame) to seven independent
operator structures in the potential, each multiplied by a coefficient
which is a linear combination of LEC's.  Specifically,
\begin{eqnarray}
v^{\rm CT2}({\bf k}, {\bf K})&=& C_1\,k^2+C_2\,K^2+
(C_3\,k^2+C_4\,K^2)\,{\bm \sigma}_1\cdot{\bm \sigma}_2 
+ i\,C_5\,\frac{{\bm \sigma}_1+{\bm \sigma}_2}{2}\cdot {\bf K}\times{\bf k} \nonumber\\
&+&C_6\,{\bm \sigma}_1\cdot{\bf k}\,\,{\bm \sigma}_2\cdot {\bf k}
+C_7\,{\bm \sigma}_1\cdot{\bf K}\,\,{\bm \sigma}_2\cdot {\bf K} \ ,
\label{eq:ct2}
\end{eqnarray}
where the $C_i$'s ($i=1,\dots,7$) are linear combinations
of the $C_i^{\prime}$'s ($i=1,\dots,14$), given by
\begin{eqnarray}
\label{eq:c1-7}
C_1 &=& C_1^{\prime} -C_3^{\prime}+C_2^{\prime}/2 \ , \nonumber \\
C_2 &=& 4\,C_1^{\prime}-4\,C_3^{\prime}-2\,C_2^{\prime} \ , \nonumber \\
C_3 &=& C_9^{\prime}+C_{12}^{\prime}/2-C_{14}^{\prime} \ , \nonumber \\
C_4 &=&  4\,C_9^{\prime}-2\,C_{12}^{\prime}+4\,C_{14}^{\prime} \ ,  \\
C_5 &=&  2\,C_5^{\prime}-4\,C_4^{\prime}-2\,C_6^{\prime} \ ,  \nonumber \\
C_6 &=& C_7^{\prime}+C_8^{\prime}+C_{10}^{\prime}/2+C_{11}^{\prime}/2-C_{13}^{\prime} \ , \nonumber \\
C_7 &=& 4\,C_7^{\prime}+4\,C_8^{\prime}
-2\,C_{10}^{\prime}-2\,C_{11}^{\prime}+4\,C_{13}^{\prime} \ ,  \nonumber
\end{eqnarray}
and, as per the comment above, one should keep in mind that any single
one of the terms proportional to $C^\prime_4$, $C^\prime_5$, and
$C^\prime_6$ ($C_7^\prime$, $C^\prime_8$, $C_{10}^\prime$, and
$C_{11}^\prime$) can be reduced to a combination of the remaining ones by
a simple redefinition of the LEC's.

The two-pion-exchange loop diagrams of panels d)-f) generate
the following contribution:
\begin{eqnarray}
\label{eq:tpe}
\!\!\!\!v^{2\pi}({\bf k})\!\!&=\!\!&v_{{\rm d}}(k)+v_{{\rm e}}(k)+v_{{\rm f}}({\bf k}) \nonumber \\
\!\!&=&\!\!\frac{g_A^2}{F_{\pi}^4}\,{\bm \tau}_1\cdot{\bm \tau}_2
\int_{\bf p} \frac{p^2-k^2}{\omega_+\,\omega_-(\omega_++\omega_-)}  
-\frac{1}{8\,F_{\pi}^4}\,{\bm \tau}_1\cdot{\bm \tau}_2
\int_{\bf p}\frac{(\omega_+-\omega_-)^2}{\omega_+\,\omega_-(\omega_++\omega_-)}\nonumber \\
\!\!&-&\!\!\frac{g_A^4}{2\,F_{\pi}^4}\,\int_{\bf p}
\frac{\omega_+^2+\omega_+\,\omega_-+\omega_-^2}{\omega_+^3\,\omega_-^3(\omega_++\omega_-)}
\left[{\bm \tau}_1\cdot{\bm \tau}_2 \,(p^2-k^2)^2
+6\,{\bm \sigma}_1\cdot({\bf k}\times{\bf p})\,{\bm \sigma}_2\cdot({\bf k}\times{\bf p})\right] \ , 
\end{eqnarray}
where $\omega_\pm=\sqrt{({\bf p}\pm {\bf k})^2+4\,m_{\pi}^2}$.
Note that recoil corrections to the reducible box diagrams have been
included in the expressions above (for a detailed discussion of this
aspect of the present study, see Section VI of Ref.~\cite{Pastore08}).
These recoil terms need also be accounted for when dealing with
loop corrections to the LO contact and OPE interactions.  The resulting
contributions, panels g) and h), are then found to be
\begin{eqnarray}
\label{eq:ctloop}
v_{\rm g}&=&\frac{4\,g_A^2}{3\,F_{\pi}^2}\, C_T\,{\bm \tau}_1\cdot{\bm \tau}_2\,
\,{\bm \sigma}_1\cdot{\bm \sigma}_2
\int_{\bf p}\, \frac{p^2}{\omega_p^3} \ ,\\
\label{eq:opeloop}
v_{\,{\rm h}}({\bf k})& =& 
-\frac{g_A^4}{3\,F_{\pi}^4}\,\frac{{\bm \tau}_1\cdot{\bm \tau}_2}{\omega_k^2}
{\bm \sigma}_1\cdot{\bf k}\,\,{\bm \sigma}_2\cdot{\bf k}
\int_{\bf p}\, \frac{p^2}{\omega_p^3} \ ,
\end{eqnarray}
The potential constructed so far is in agreement with that obtained
by Epelbaum {\it et al.} in Ref.~\cite{Epelbaum98} by the method of
unitary transformations, but for an overall factor of +8/3 rather than --1/3
in Eq.~(\ref{eq:opeloop}).

Lastly, we consider the diagram illustrated in panel i),
which has both reducible and irreducible parts.
The former describe interactions involving
``dressed nucleons''.  We do not take into account recoil
corrections arising from the pion emitted and reabsorbed by the
same nucleon.  Hence, for diagram i) we retain
the irreducible part only, and obtain
\begin{eqnarray}
\label{eq:ctloop1}
v_{{\rm i}}&=&\frac{g_A^2}{F_{\pi}^2}\,
(3\,C_S-C_T\,{\bm \sigma}_1\cdot{\bm \sigma}_2)\int_{\bf p}\,\frac{p^2}{\omega_p^3}\ .
\end{eqnarray}
Again, this approach leads to a result which differs from that reported
in Ref.~\cite{Epelbaum98} for this
diagram, specifically the term proportional to $C_S$
in Eq.~(\ref{eq:ctloop1}) is absent, while that proportional
to $C_T$ is multiplied by
$-4\, (g_A^2/F_\pi^2) $ rather than $- (g_A^2/F_\pi^2)$.  However,
as it will become clear in the next section, these differences---for diagrams h) and i)---do
not affect the definition of the renormalized potential,
since they only lead to differences in the renormalization of
the LEC's $C_S$, $C_T$, and $g_A$.

\subsection{Renormalization}
\label{sec:pot_ren}

The potential defined in the previous section contains ultraviolet
divergencies which need to be removed by a proper renormalization
procedure.  In order to isolate these divergencies, the kernels of
the N$^2$LO contributions have been evaluated using dimensional
regularization, and the relevant integration formulae are listed in
Appendix~\ref{app:dimensional}.  Here we sketch the regularization
procedure of the various contributions, and give the final expression
for the renormalized {\it NN} potential.

As an example, we discuss, in some detail, the regularization of
the two-pion-exchange contribution of Eq.~(\ref{eq:tpe}).  In terms of
the kernels $L(k)$, $I^{(2n)}(k)$ and $J^{(2n)}(k)$ defined in
Appendix \ref{app:dimensional}, it reads as
\begin{eqnarray}
v^{2\pi}({\bf k})=&-&\frac{1}{8\,F_{\pi}^4}\,{\bm \tau}_1\cdot{\bm \tau}_2\,
\Bigg[L(k)-8\,g_A^2\Big[I^{(2)}(k)-k^2I^{(0)}(k)\Big]
+4\,g_A^4\Big[J^{(4)}(k)\nonumber\\
&-  & 2\, k^2J^{(2)}(k)+k^4J^{(0)}(k)\Big]\Bigg]
-\frac{3\,g_A^4}
{F_{\pi}^4}\,({\bm \sigma}_1\times{\bf k})_i({\bm \sigma}_2\times{\bf k})_j \,J_{ij}^{(2)}(k) \ .
\end{eqnarray} 
By inserting the explicit expressions of these kernels in the previous equation, we obtain
\begin{equation}
\label{eq:tpe1}
v^{2\pi}({\bf k})=\overline{v}^{2\pi}({\bf k})+{\bm \tau}_1\cdot{\bm \tau}_2\, P_2(k)
+\left( k^2\,{\bm \sigma}_1\cdot{\bm \sigma}_2-{\bm \sigma}_1 \cdot{\bf k}\,\,{\bm \sigma}_2\cdot{\bf k}\right)\, P_0 \ ,
\end{equation}
where the renormalized (finite) part of the two-pion-exchange potential,
denoted by $\overline{v}^{2\pi}({\bf k})$, is given by
\begin{eqnarray}
\overline{v}^{2\pi}({\bf k})\!\!&=&\!\!\frac{1}
{48 \pi^2\,F_{\pi}^4}{\bm \tau}_1\cdot{\bm \tau}_2\,
G(k) \left[4 m_{\pi}^2(1+4 g_A^2-5 g_A^4)+k^2(1+10 g_A^2 - 23 g_A^4)-\frac{48\,g_A^4 m^4_\pi}
{4\,  m^2_\pi+k^2}\right] \nonumber \\
&+&\frac{3\,g_A^4}{8\pi^2\,F_{\pi}^4}\,G(k) \left( k^2\,{\bm \sigma}_1\cdot{\bm \sigma}_2-
{\bm \sigma}_1 \cdot{\bf k}\, {\bm \sigma}_2\cdot{\bf k}\right) \ ,
\label{eq:tper}
\end{eqnarray}
with
\begin{equation}
G(k)=\frac{\sqrt{4\,m_{\pi}^2+k^2}}{k}\ln \frac{\sqrt{4\,m_{\pi}^2+k^2}+k}{\sqrt{4\,m_{\pi}^2+k^2}-k} \ ,
\label{eq:loopf}
\end{equation}
where the loop function $G(k)$ defined here differs
by a factor two from that given in Ref.~\cite{Epelbaum98}.

The divergencies are lumped into the polynomials $P_2(k)$ (of order two) and constant $P_0$:
\begin{eqnarray}
P_2(k)\!\! =\!\!&-&\!\!\frac{1}{24 \pi^2\,F_{\pi}^4}
  \Bigg[ m_{\pi}^2 \Big[ 4 +22 g_A^2-29 g_A^4-9 g_A^2( 2 -5g_A^2)
\Big(-\frac{2}{\epsilon}+\gamma-\ln \pi + \ln \frac{m_{\pi}^2}{\mu^2}\Big) \Big]  \nonumber\\
&+&\!\!\frac{4}{3} k^2 \Big[ 1+7 g_A^2-9 g_A^4-\frac{3}{8}\,(1+10 g_A^2 - 23 g_A^4)
\Big(-\frac{2}{\epsilon}+\gamma-\ln \pi + \ln \frac{m_{\pi}^2}{\mu^2}\Big)\Big] \Bigg] \ ,
\end{eqnarray}
\begin{equation}
P_0 =\frac{3\,g_A^4}{8\pi^2\,F_{\pi}^4} 
\left( -\frac{2}{\epsilon}+\gamma-\ln \pi + \ln \frac{m_{\pi}^2}{\mu^2}-\frac{4}{3}\right) \  .
\end{equation}
where the parameter $\epsilon \rightarrow 0^+$, 
$\gamma$ is the Euler-Mascheroni constant, and $\mu$ is the
renormalization scale brought in by the dimensional regularization procedure.
The isospin structure ${\bm \tau}_1\cdot{\bm \tau}_2$  multiplying the polynomial
$P_2(k)$ can be reduced by Fierz rearrangement so as to match structures occurring
in the LO $v^{\rm CT0}$ and N$^2$LO $v^{\rm CT2}({\bf k})$ contact
contributions.  Indeed, because of the antisymmetry of
two-nucleon states,
\begin{eqnarray}
{\bm \tau}_1\cdot {\bm \tau}_2 &=&- 2 - {\bm \sigma}_1 \cdot {\bm
  \sigma}_2 \ ,\\
 {\bm \tau}_1\cdot {\bm \tau}_2 \, k^2&=&-  4\, (1 + {\bm \sigma}_1 \cdot
{\bm \sigma}_2) \, K^2 -  k^2 \ .
\end{eqnarray}
It is then seen that the terms in $P_0$ and $P_2(k)$ renormalize $C_S$, $C_T$, $C_1$,
$C_2$, $C_4$ and $C_6$. For example, the last term of Eq.~(\ref{eq:tpe1}) is
absorbed by the redefinition,
\begin{equation}
\label{eq:c6ren}
C_6=\overline{C}_6 + \frac{3 g_A^4}{8 \pi^2 F_\pi^4}\, \mu^{-\epsilon}\, \left(
-\frac{2}{\epsilon} + \gamma - \ln \pi + \ln \frac{m_\pi^2}{\mu^2}
-\frac{4}{3}\right),
\end{equation}
where the factor $\mu^{-\epsilon}$ is needed because the mass dimension of
the LEC $C_6$ is $d-7$ in $d$ space dimensions.  Note that the
renormalized $\overline{C}_6$ remains $\mu$-independent, as becomes obvious by
taking the logarithmic derivative with respect to $\mu$ and neglecting
$O(\epsilon)$ terms.  For ease of notation, we will omit the overline and
tacitly imply that the LECs have been properly renormalized.

The contributions in Eqs.~(\ref{eq:ctloop}), (\ref{eq:ctloop1}), and (\ref{eq:opeloop})
lead to further renormalization of the  LEC's $C_S$ and $C_T$,
as well as the axial coupling constant $g_A$ entering the LO OPE:
\begin{eqnarray}
\label{eq:vgi}
 v_{\rm g}+v_{\rm i}&=&
\frac{4\,g_A^2}{3\,F_{\pi}^2}\, C_T\, {\bm \tau}_1\cdot{\bm \tau}_2
\,{\bm \sigma}_1\cdot{\bm \sigma}_2\,M^{(3)}
+\frac{g_A^2}{F_{\pi}^2}(3\, C_S-C_T\,
{\bm \sigma}_1\cdot{\bm \sigma}_2) \, M^{(3)}\ , \\
 v_{\,{\rm h}}({\bf k})&=&
-\frac{g_A^4}{3\,F_{\pi}^4}\,{\bm \tau}_1\cdot{\bm \tau}_2\,
\frac{{\bm \sigma}_1\cdot{\bf k}\,\,{\bm\sigma}_2\cdot{\bf k} }{\omega_k^2}\, M^{(3)}\ ,
\label{eq:vgh}
\end{eqnarray}
where the constants $M^{(n)}$ are listed in
Appendix~\ref{app:dimensional}.  The complete $N\!N$ potential up to N$^2$LO included
is then given as
\begin{equation}
v({\bf k},{\bf K})=\overline{v}^{\rm CT0}+\overline{v}^\pi({\bf k})+\overline{v}^{\rm CT2}({\bf k},{\bf K})
+\overline{v}^{2\pi}({\bf k}) \ ,
\label{eq:vn2lo}
\end{equation}
where $\overline{v}^{\rm CT0}$, $\overline{v}^\pi$,
$\overline{v}^{\rm CT2}$, and $\overline{v}^{2\pi}$ are defined in Eqs.~(\ref{eq:ct0}),
(\ref{eq:ope}), (\ref{eq:ct2}), and~(\ref{eq:tper}), respectively,
and the overline indicates that the LEC's $g_A$ and some of the $C_i^\prime$ have
been renormalized.
\section{Electromagnetic currents}
\label{sec:currents}

\begin{figure}[bthp]
\includegraphics[width=4in]{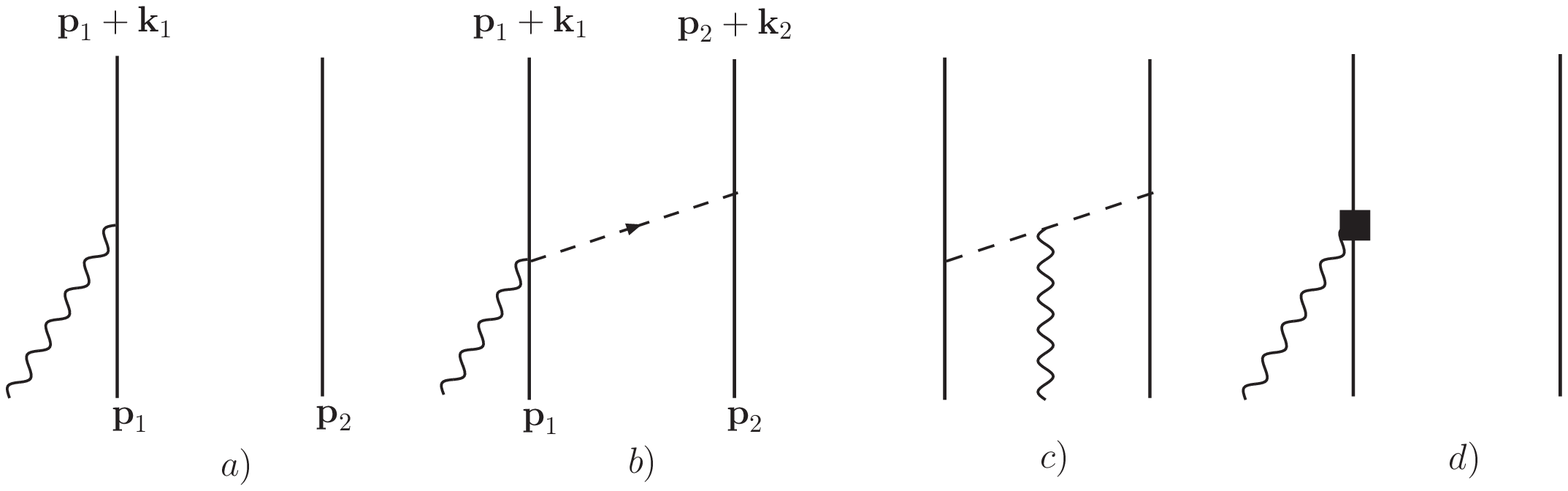}
\caption{Diagrams illustrating one- and two-body currents entering at LO ($e\, Q^{-2}$),
panel a), NLO ($e\, Q^{-1}$), panels b) and c),
and N$^2$LO  ($e\, Q^{\,0}$), panel d).  Nucleons, pions, and photons are denoted by solid, dashed, and wavy lines,
respectively.  The square represents the $(Q/m_N)^2$ relativistic correction to the
LO one-body current.  Only one among the possible time orderings is shown for the NLO
currents.}
\label{fig:fig2}
\end{figure}

In this section we construct the electromagnetic current
operator for a two-nucleon system in $\chi$EFT.  In the
power-counting scheme of Ref.~\cite{Pastore08}, the
LO term results from the coupling of the external photon
field to the individual nucleons, and is counted as $e\,Q^{-2}$,
where a factor $e\,Q$ is from the $\gamma NN$ vertex,
and a factor $Q^{-3}$ follows from the momentum
$\delta$-function implicit in this type of disconnected
diagrams, see panel a) of Fig.~\ref{fig:fig2}.  Higher
order terms are suppressed by additional powers
of $Q$, and formal expressions up to N$^3$LO, {\it i.e.}~$e\, Q$,
have been derived in Ref.~\cite{Pastore08}.
In this section, we proceed to regularize the loop integrals
entering these N$^3$LO currents, and to derive the corresponding
finite parts.

At this order, we distinguish among four classes of contributions:
i) currents generated by minimal substitution in the four-nucleon
contact interactions involving two gradients of the nucleons' fields,
as well as by non-minimal couplings (these were not considered
in Ref.~\cite{Pastore08}); ii) two-pion exchange currents at one loop;
iii) one-loop corrections to tree-level currents; and iv) $(Q/M)^2$ relativistic
corrections to the NLO currents resulting from
the non-relativistic reduction of the vertices.  The latter are
neglected in the present work.

For completeness, we report below the expressions up to N$^2$LO
derived in Ref.~\cite{Pastore08}, and shown in Fig.~\ref{fig:fig2}.
As emphasized earlier, in the present study we do not
explicitly include $\Delta$-isobar degrees of freedom.  In the
following, the momenta are defined as
\begin{equation}
{\bf k}_i = {\bf p}_i^\prime -{\bf p}_i \ , \qquad {\bf K}_i =({\bf p}_i^\prime +{\bf p}_i)/2 \ ,
\label{eq:defs}
\end{equation}
where ${\bf p}_i$ and ${\bf p}_i^\prime$ are the initial and final momenta of nucleon $i$.

The LO contribution of panel a) in Fig.~\ref{fig:fig2} is
\begin{equation}
{\bf j}_{\rm a}^{\rm LO}=\frac{e}{2\, m_N}
\Big[ 2\, e_{N,1} \, {\bf K}_1 +i\,\mu_{N,1}\, {\bm \sigma}_1\times {\bf q }\Big] +
 1 \rightleftharpoons 2\ ,
 \label{eq:jlo}
\end{equation}
where ${\bf q}$ is the photon momentum, ${\bf q}={\bf k}_i$, and
\begin{equation}
e_N = (1+\tau_z)/2 \ , \qquad 
\kappa_N =  (\kappa_S+ \kappa_V \, \tau_z)/2 \ , \qquad \mu_N = e_N+\kappa_N \ ,
\label{eq:ekm}
\end{equation}
$\kappa_S$ and $\kappa_V$ being
the isoscalar and isovector combinations
of the anomalous magnetic moments of
the proton and neutron ($\kappa_S =-0.12$ $\mu_N$
and $\kappa_V =3.706$ $\mu_N$).  Loop corrections
to the one-body current above, occurring at NLO and N$^2$LO,
are absorbed into $\kappa_S$ and $\kappa_V$.
The NLO seagull and pion-in-flight contributions,
represented in panels b) and c), are:
\begin{eqnarray}
 {\bf j}^{\rm NLO}_{\rm b}&=& -i\, e\frac{g^2_A}{F^2_\pi}\,
 ({\bm \tau}_1 \times {\bm \tau}_2)_z \,\,
 {\bm \sigma}_1 \, \frac{{\bm \sigma}_2\cdot {\bf k}_2}{\omega^2_{k_2}}+ 1 \rightleftharpoons 2 \ , 
 \label{eq:nlo1} \\
  {\bf j}^{\rm NLO}_{\rm  c}&=& i\, e\frac{g^2_A}{F^2_\pi}\,
 ({\bm \tau}_1 \times {\bm \tau}_2)_z 
\frac{{\bf k}_1-{\bf k}_2}{\omega^2_{k_1}\,\omega^2_{k_2}}
  {\bm \sigma}_1\cdot {\bf k}_1 \,\, {\bm \sigma}_2\cdot {\bf k}_2 \ ,
\label{eq:nlo2}
\end{eqnarray}
where the momenta transferred to nucleons 1 and 2
add up to ${\bf q}$, ${\bf k}_1+{\bf k}_2={\bf q}$.  
Lastly, the N$^2$LO (relativistic) correction to the LO current, represented in
panel d), reads:
\begin{eqnarray}
 {\bf j}_{\rm d}^{\rm N^2LO}=&-&\frac{e}{8 \, m_N^3}
 e_{N,1}\, \Bigg[ 
2\, \left( K_1^2 +q^2/4 \right) 
 \left( 2\, {\bf K}_1+i\, {\bm \sigma}_1\times {\bf q } \right)
+ {\bf K}_1\cdot {\bf q}\,
 \left({\bf q} +2\, i\, {\bm \sigma}_1\times {\bf K }_1 \right)\Bigg]
 \nonumber \\ 
 &-& \frac{i\,e}{8 \, m_N^3}
 \kappa_{N,1}\, \Bigg[ {\bf K}_1\cdot {\bf q}\, 
 \left( 4\, {\bm \sigma}_1\times {\bf K}_1 -i\, {\bf q}\right) 
 - \left(  2\, i\, {\bf K}_1 -{\bm \sigma}_1\times {\bf q} \right)\, q^2/2 \nonumber \\
 && \qquad\qquad \qquad +2\, \left({\bf K}_1\times {\bf q}\right)
 \, {\bm \sigma}_1\cdot {\bf K}_1
 \Bigg] + 1 \rightleftharpoons 2  \ .
\label{eq:j1rc}
\end{eqnarray}

In addition to the classes mentioned earlier, there
are N$^3$LO contributions~\cite{Phillips09} involving
the standard $\pi NN$ vertex on one nucleon, and $\gamma\pi NN$
vertices of order $e\, Q^2$ on the other nucleon, derived from the following
interaction Hamiltonian~\cite{Fettes98}
\begin{eqnarray}
H^{(2)}_{\gamma\pi NN}&=&\frac{e}{F_\pi} \int {\rm d}{\bf x}\, N^\dagger({\bf x})
\Big[ d_8^{\,\prime}\, \nabla \pi_z({\bf x})+d_9^{\,\prime}\, 
\tau_a\, \nabla \pi_a({\bf x}) \nonumber \\
 &-& d_{21}^{\, \prime}\, \epsilon_{zab}\, \tau_a \,{\bm \sigma} \times \nabla \pi_b({\bf x})
\Big]N({\bf x}) \cdot \nabla \times {\bf A}({\bf x}) \ ,
\end{eqnarray}
where the notation and conventions of Ref.~\cite{Pastore08}
have been adopted for the nucleon ($N$), pion ($\pi_a$), and
photon (${\bf A}$) fields, and $d_8^{\, \prime}$,
$d_9^{\, \prime}$, and $d_{21}^{\, \prime}$ are related to the original
couplings given by Fettes {\it et al.}~\cite{Fettes98} via
$d_8^{\, \prime}=8\,[d_8+g_A/(64 \, m_N^2)]$ and similarly
for $d_9^{\, \prime}$, and $d_{21}^{\, \prime}=2\,d_{21}+d_{22}$.
The resulting two-body current is given by
\begin{equation}
{\bf j}^{\rm N^3LO}_{\rm tree}= i\, e\frac{g_A}{F_\pi^2}\Bigg[
\left(d_8^{\, \prime}\, \tau_{2,z} + 
d_9^{\, \prime}\, {\bm \tau}_1 \cdot {\bm \tau}_2 \right){\bf k}_2 
-d_{21}^{\, \prime} ({\bm \tau}_1\times{\bm \tau}_2)_z {\bm \sigma}_1\times {\bf k}_2
\Bigg] \times {\bf q}\,
\frac{{\bm \sigma}_2\cdot {\bf k}_2}{\omega_{k_2}^2}
+ 1 \rightleftharpoons 2 \ ,
\end{equation}
and in principle the unknown LEC's $d_8^{\, \prime}$,
$d_9^{\, \prime}$, and $d_{21}^{\, \prime}$ could be determined
by pion photoproduction data on a single nucleon or nuclear data
(as discussed in Sec.~\ref{sec:intro}).  The isovector
part of ${\bf j}^{\rm N^3LO}_{\rm tree}$ has the same structure as the current
involving $N$-$\Delta$ excitation~\cite{Pastore08}, to which
it reduces if the following identifications are made:
$d_{21}^{\, \prime}/d_8^{\, \prime}=1/4$, and
$d_8^{\, \prime} = 4\, \mu^* h_A /(9\, m_N \,\Delta)$,
where $h_A$ is the $\pi N \Delta$ coupling constant, $\mu^*$
is the $N\Delta$-transition magnetic moment, and $\Delta$ is
the $\Delta$-$N$ mass difference, $\Delta=m_\Delta-m_N$.

Configuration-space representations of the current operators follow from
\begin{eqnarray}
{\bf j}^{(1)}({\bf q}) &=& \int_{{\bf k}_1} \int_{{\bf K}_1} 
{\rm e}^{i {\bf k}_1\cdot ({\bf r}^\prime_1+{\bf r}_1)/2}\,
{\rm e}^{i {\bf K}_1\cdot ({\bf r}^\prime_1-{\bf r}_1)}\,\,
\overline{\delta}({\bf k}_1-{\bf q})\,\,
{\bf j}^{(1)}({\bf k}_1,{\bf K}_1) \ , 
\label{eq:j1rs}\\
 {\bf j}^{(2)}({\bf q}) &=& \int_{{\bf k}_1}\int_{{\bf k}_2}
{\rm e}^{i {\bf k}_1\cdot {\bf r}_1}\,
{\rm e}^{i {\bf k}_2\cdot {\bf r}_2}\,\,
\overline{\delta}({\bf k}_1+{\bf k}_2-{\bf q})\,\,
{\bf j}^{(2)}({\bf k}_1,{\bf k}_2) \nonumber\\ 
&=&{\rm e}^{i\, {\bf q}\cdot {\bf R} } \int_{\bf k} {\rm e}^{i\, {\bf k}\cdot {\bf r} }\,
{\bf j}^{(2)}({\bf q},{\bf k})
\label{eq:j2rs} \ ,
\end{eqnarray}
where ${\bf j}^{(1)}$ or ${\bf j}^{(2)}$ denote any
one-body or two-body operators, respectively, and
$\overline{\delta}(\dots) \equiv (2\pi)^3\delta(\dots) $.
Note that ${\bf K}_i \rightarrow -i \nabla_i^\prime 
\delta({\bf r}_i^\prime-{\bf r}_i)$, {\it i.e.}
the configuration-space representation of the momentum operator,
and in the second line of Eq.~(\ref{eq:j2rs}) ${\bf R}$ and ${\bf r}$
are the center-of-mass and relative positions of the two nucleons.
\subsection{N$^3$LO currents: terms from four-nucleon contact interactions}
\label{sec:ctcnt}
The N$^3$LO currents obtained by minimal substitution in the
contact interactions involving two gradients of the nucleons' fields
have been constructed in Ref.~\cite{Pastore08}, and are reported
below for reference:
\begin{eqnarray}
\label{eq:counter}
{\bf j}_{\rm CT\gamma}^{\rm N^3LO}&=& -e \, e_1\bigg[
2\, \left(2\, C^{\prime}_{1}-C_{2}^{\prime}\right)\, {\bf K}_{2}+4\, C^{\prime}_{3}\,{\bf K}_{1}
+i\,C_{4}^{\prime}\, \left({\bm \sigma}_{1}+
{\bm \sigma}_{2}\right)\times{\bf k}_{2}
 + i \,C^{\prime}_{5}\, {\bm \sigma}_1\times{\bf k}_1 \nonumber \\
&-&i \,C^{\prime}_{6}\,
{\bm \sigma}_2\times{\bf k}_1 + 2\, \left(2\, C^{\prime}_7-
C^{\prime}_{10}\right)({\bf K}_{2}\cdot{\bm \sigma}_2)\, {\bm \sigma}_1 + 
  2\, \left(2\, C^{\prime}_8-C^{\prime}_{11}\right)({\bf K}_{2}
 \cdot{\bm \sigma}_1)\, {\bm \sigma}_2  \nonumber \\
&-&2\, C^{\prime}_{13}\, \left[({\bf K}_1\cdot{\bm \sigma}_1)\,
{\bm \sigma}_2+ ({\bf K}_1\cdot {\bm \sigma}_2)\, {\bm \sigma}_1 \right] 
+ 2\, \left(2\, C^{\prime}_{9}-C^{\prime}_{12}\right){\bf K}_2\,
({\bm \sigma}_1\cdot{\bm \sigma}_2)\nonumber \\
&-&4\, C^{\prime}_{14}\,{\bf K}_1\,
({\bm \sigma}_1\cdot{\bm \sigma}_2) \bigg] + 1 \rightleftharpoons 2 \ .
\end{eqnarray}
In addition to these, there are contributions due to
non-minimal couplings, as derived in Appendix~\ref{app:nmcounterterms},
\begin{equation}
\label{eq:nmcounter}
{\bf j}_{\rm CT\gamma nm}^{\rm N^3LO}= - i\,e \bigg[ C_{15}^\prime\,
{\bm \sigma}_1  + C_{16}^\prime\,
  (\tau_{1,z} - \tau_{2,z})\,{\bm \sigma}_1  \bigg]\times {\bf q}  
+ 1 \rightleftharpoons 2 \ .
\end{equation}
\begin{figure}
\includegraphics[width=3.5in]{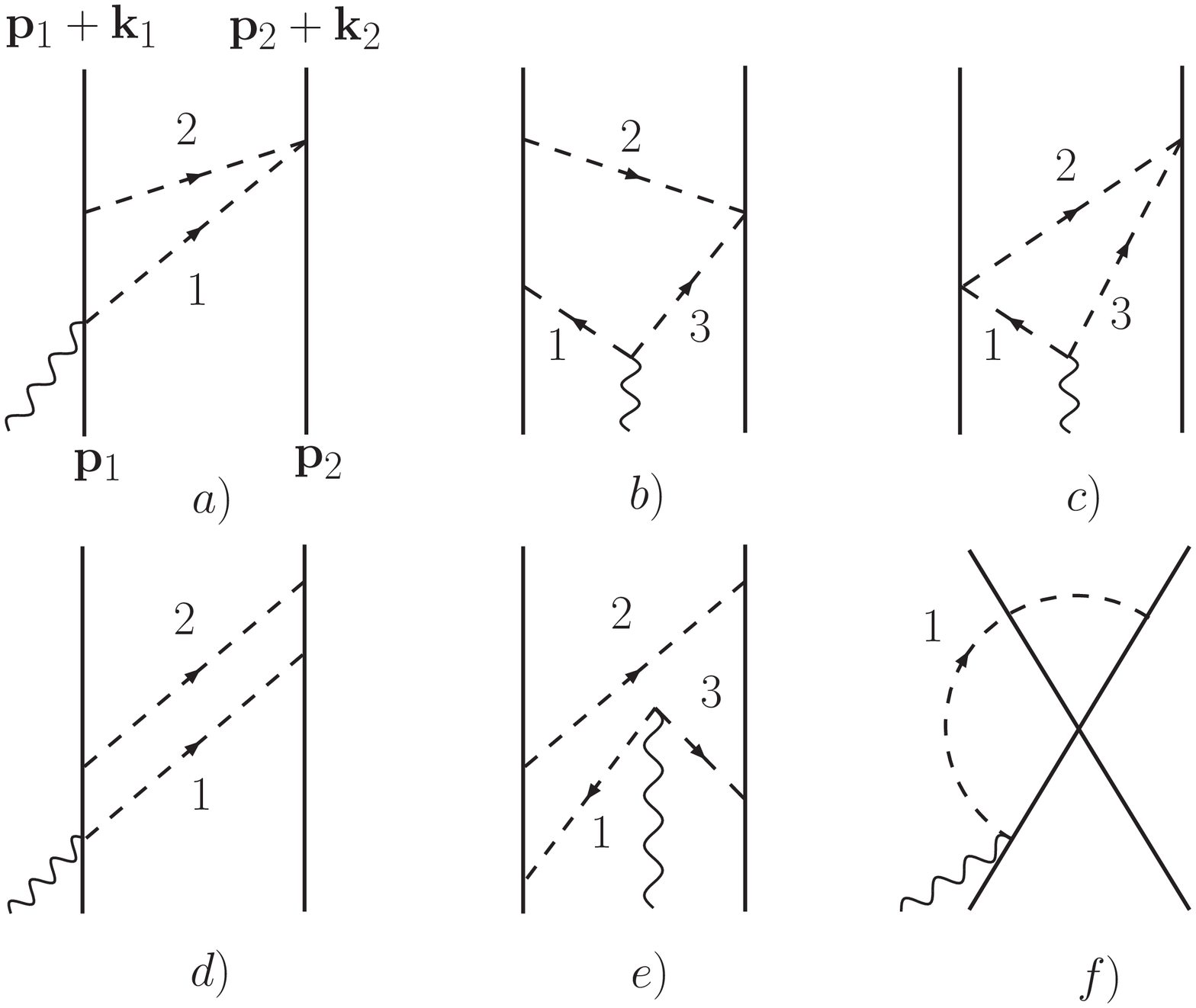}\\
\includegraphics[width=3.5in]{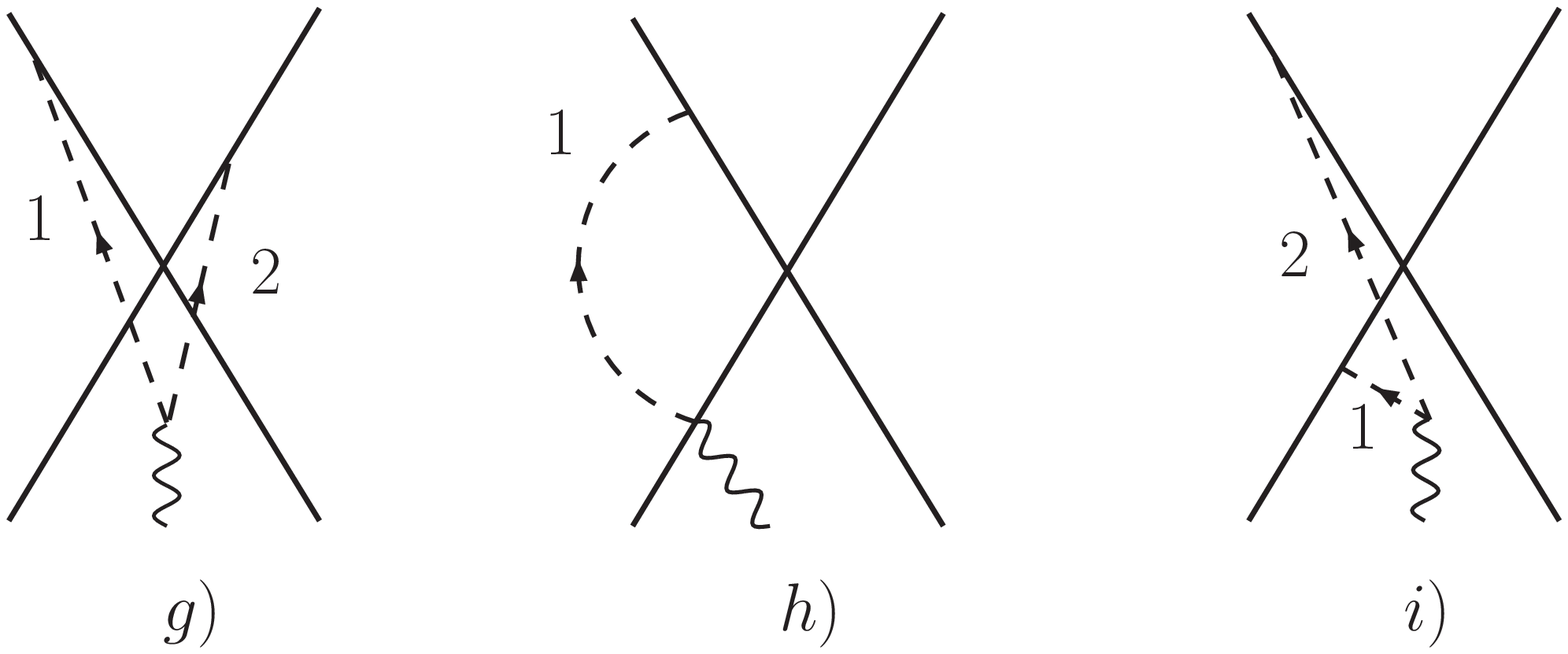}
\caption{Diagrams illustrating one-loop two-body currents entering at N$^3$LO
($e\, Q$), notation as in Fig.~\protect\ref{fig:fig2}.
Only one among the possible time orderings is shown for each contribution.}
\label{fig:fig9}
\end{figure}
\subsection{N$^3$LO currents: one-loop corrections}
\label{sec:llcc}

Loop corrections entering at N$^3$LO
have been derived in Ref.~\cite{Pastore08}, and their formal
expressions, corrected from a number of typographical errors,
are listed, for reference, in Appendix~\ref{app:cntold} of the
present paper.  In Ref.~\cite{Pastore08}, it was also shown
that the one-loop currents satisfy the continuity equation with
the two-pion-exchange potential of Sec.~\ref{sec:pot_formulae}.
Here we discuss their renormalization.  We start off by considering the currents
 (involving one and two pions) illustrated in panels a), d), f), g), h), and i) of
Fig.~\ref{fig:fig9}.  Those in panels b), c), and e) (involving three pions)
are discussed in Sec.~\ref{sec:mum} and Appendix~\ref{app:muloop}, since
for them we only derive the magnetic dipole operators.  In terms of the
kernels defined in Appendix~\ref{app:dimensional}, we obtain:
\begin{eqnarray}
 {\bf j}^{\rm N^3LO}_{\rm a}&=&-\,i\,e\,\frac{g_A^2}
{F_{\pi}^4}\,I^{(0)}(k_2) \Big[2\,\tau_{2,z}\,{\bm \sigma_1}\times {\bf k}_2
+({\bm \tau}_1\times{\bm \tau}_2)_z\, {\bf k}_2\Big]+ 1 \rightleftharpoons 2 \ , \\
{\bf j}^{\rm N^3LO}_{\rm d}&=&\, -i\,e\, \frac{g_A^4}{F_{\pi}^4}\Bigg[\left[ k_2^2\,  J^{(0)}(k_2)-J^{(2)}(k_2)\right]
\Big[2\, \tau_{2,z}\,{\bm \sigma_1}\times {\bf k}_2 +
({\bm \tau}_1\times{\bm \tau}_2)_z\,{\bf k}_2\Big]  \nonumber \\
&+& 4\, \tau_{1,z} \, J^{(2)}_{ij}(k_2)\, \left({\bm \sigma}_2\times {\bf k}_2 \right)_j\Bigg] + 
1 \rightleftharpoons 2  \ ,\\
{\bf j}^{\rm N^3LO}_{\rm g}&=&-2\, i\,e\frac{g_A^2}{F_{\pi}^2}\, C_T\, ({\bm \tau}_1\times{\bm \tau}_2)_z\, J^{(2)}_{ij}(q)\,
{\sigma}_{1,j}\,\,{\bm \sigma}_2\cdot{\bf q}+ 1 \rightleftharpoons 2 \ , \\ 
 {\bf j}^{\rm N^3LO}_{\rm i}&=&-2\, i\,e\frac{g_A^2}{F_{\pi}^2} 
\tau_{1,z} \,J^{(2)}_{ij}(q)\,
 \Big[C_S\, ({\bm \sigma}_1\times{\bf q})_j -
C_T\,({\bm \sigma}_2\times{\bf q})_j\Big] + 1 \rightleftharpoons 2 \ ,
\end{eqnarray}
and the currents in panels f) and h) vanish in the static limit~\cite{Pastore08}. Insertion of the finite parts
of the various kernels in the expressions above then gives
\begin{eqnarray}
 \overline{{\bf j}}^{\rm N^3LO}_{\rm a}&=&
\,i\,e\,\frac{g_A^2}{8\pi^2F_{\pi}^4}\,G(k_2) \Big[2\, \tau_{2,z}\,{\bm \sigma_1}\times {\bf k}_2
+({\bm \tau}_1\times{\bm \tau}_2)_z\, {\bf k}_2\Big]+ 1 \rightleftharpoons 2 \ , \\
\overline{{\bf j}}^{\rm N^3LO}_{\rm d}&=&\, -i\,e\, \frac{g_A^4}{8\,\pi^2\,F_{\pi}^4}G(k_2)\, \Bigg[
\Bigg(3-\frac{4\,m_{\pi}^2}{4\,m_{\pi}^2+k_2^2} \Bigg)
\Big[ 2\, \tau_{2,z}\,{\bm \sigma_1}\times {\bf k}_2 +
({\bm \tau}_1\times{\bm \tau}_2)_z\,{\bf k}_2\Big] \nonumber \\
&-& 4\, \tau_{1,z}\,{\bm \sigma}_{2}\times {\bf k}_{2}\Bigg] + 1 \rightleftharpoons 2 \ ,\\
\overline{{\bf j}}^{\rm N^3LO}_{\rm g}&=& i\,e\frac{g_A^2\,C_T}{4\,\pi^2\,F_{\pi}^2}
 ({\bm \tau}_1\times{\bm \tau}_2)_z\,
G(q)\, \,{\bm \sigma}_{1}\,{\bm \sigma}_2\cdot{\bf q}+ 1 \rightleftharpoons 2 \  , \\ 
\overline{{\bf j}}^{\rm N^3LO}_{\rm i}&=& i\, e\,
\frac{g_A^2}{4\pi^2\,F_{\pi}^2} \tau_{1,z} \, G(q)\,
(C_S\, {\bm \sigma}_1\times{\bf q}-
C_T\,{\bm \sigma}_2\times{\bf q}) + 1 \rightleftharpoons 2 \ ,
\end{eqnarray}
and the loop function $G$ is defined in Eq.~(\ref{eq:loopf}).
The divergent parts of the kernels lead to renormalization of some of the
LEC's $C_i^\prime$.  They are given by
\begin{eqnarray}
{\bf j}^{\rm N^3LO}_{\infty,{\rm a}}&=&i\,e\,\frac{g_A^2}{8\pi^2F_{\pi}^4}
\left(\frac{2}{\epsilon}+\ldots \right)
\Big[-2\, \tau_{2,z}\,{\bm \sigma_1}\times {\bf k}_2
-({\bm \tau}_1\times{\bm \tau}_2)_z\, {\bf k}_2\Big]+ 1
\rightleftharpoons 2 \ ,\\
{\bf j}^{\rm N^3LO}_{\infty,{\rm b}}&=&i\,e\,\frac{g_A^2}{8\pi^2F_{\pi}^4}
\left(\frac{2}{\epsilon}+\ldots \right)
\Bigg[2\, \tau_{2,z}\,{\bm \sigma_1}\times ({\bf k}_2-{\bf q})
-\frac{2}{3} ({\bm \tau}_1\times{\bm \tau}_2)_z\, {\bf k}_2\Bigg]+
1\rightleftharpoons 2 \ ,\\
{\bf j}^{\rm N^3LO}_{\infty,{\rm c}}&=&i\, e\frac{1}{48\pi^2F_{\pi}^4}
\left(\frac{2}{\epsilon}+\ldots \right) ({\bm \tau}_1\times{\bm \tau}_2)_z\,
({\bf k}_1-{\bf k}_2)  \ ,\\
{\bf j}^{\rm N^3LO}_{\infty,{\rm d}}&=&i\,e\,\frac{g_A^4}{8\pi^2F_{\pi}^4}
\left(\frac{2}{\epsilon}+\ldots \right)
\Big[ \tau_{2,z}\,{\bm \sigma}_1\times(6\, {\bf k}_2 - 4\, {\bf k}_1)  + 3  ({\bm \tau}_1\times{\bm \tau}_2)_z\, {\bf k}_2\Big]+ 1\rightleftharpoons 2 \ ,\\
{\bf j}^{\rm N^3LO}_{\infty,{\rm e}}&=&i\,e\,\frac{g_A^4}{8\pi^2F_{\pi}^4}
\left(\frac{2}{\epsilon}+\ldots \right)
\Bigg[10\, \tau_{2,z}\,{\bm \sigma}_1\times {\bf k}_1 + \frac{5}{6}  ({\bm \tau}_1\times{\bm \tau}_2)_z\, {\bf k}_2\Bigg]+ 1\rightleftharpoons 2 \ ,\\
{\bf j}^{\rm N^3LO}_{\infty,{\rm
    g}}&=&i\,e\,\frac{g_A^2}{4\pi^2F_{\pi}^2}
\left(\frac{2}{\epsilon}+\ldots \right)({\bm \tau}_1\times{\bm \tau}_2)_z\, C_T\,
\Big[{\bm \sigma}_2 \,\,{\bm \sigma}_1\cdot {\bf q} - {\bm \sigma}_1 \,\,
  {\bm \sigma}_2 \cdot {\bf q}\Big] \ ,\\
{\bf j}^{\rm N^3LO}_{\infty,{\rm i}}&=&i\,e\,\frac{g_A^2}{4\pi^2F_{\pi}^2}
\left(\frac{2}{\epsilon}+\ldots \right)
\tau_{1,z} \Big[ C_T\, {\bm \sigma}_2 \times {\bf q} - C_S \,{\bm
    \sigma}_1 \times {\bf q} ] + 1\rightleftharpoons 2 \ ,
\end{eqnarray}
where the dots denote finite contributions depending on the
renormalization point. When combined together, all these divergencies can be absorbed by the
renormalization of the $C^\prime_i$, which is not the case for the
individual contributions. For instance, taking into account the
antisymmetry properties of nucleons,
\begin{eqnarray}
(\tau_{2,z}\, {\bm \sigma}_1 + \tau_{1,z}\, {\bm \sigma}_2) \times {\bf q}
  &=& -(\tau_{1,z}\, {\bm \sigma}_1 + \tau_{2,z}\, {\bm \sigma}_2) \times
  {\bf q}\nonumber\\
&=& \frac{1}{2} ({\bm \tau}_1\times{\bm \tau}_2)_z
\Big[{\bm \sigma}_1 \,\, {\bm \sigma}_2\cdot {\bf q} - {\bm \sigma}_2 \,\,
  {\bm \sigma}_1 \cdot {\bf q} \Big] \nonumber \\
&=&-\frac{1}{2} (\tau_{1,z} - \tau_{2,z})\, ( {\bm \sigma}_1 - {\bm
    \sigma}_2) \times {\bf q}\ ,
\label{eq:anty}
\end{eqnarray}
leading to renormalization of $C_{16}^\prime$, and
\begin{equation}
({\bm \tau}_1\times{\bm \tau}_2)_z\,( {\bf k}_2 - {\bf k}_1) = -2\, i\,
  e_1\, ( 1 + {\bm \sigma}_1 \cdot {\bm \sigma}_2) 
({\bf K}_1 - {\bf K}_2) + 1\rightleftharpoons 2\ , 
\end{equation}
leading to renormalization of $C_{3}^\prime$, $C_{14}^\prime$, $(2\,
C_1^\prime - C_2^\prime)$ and  $(2\,
C_9^\prime - C_{12}^\prime)$.

\subsection{N$^3$LO currents: one-loop corrections to tree-level currents}
\label{sec:lp1}

Contributions in this class are illustrated by the diagrams in
Figs.~\ref{fig:fig5} and~\ref{fig:fig5n}.  After including all
possible time orderings, we find for those in Fig.~\ref{fig:fig5}:
\begin{figure}[bthp]
\includegraphics[width=4.5in]{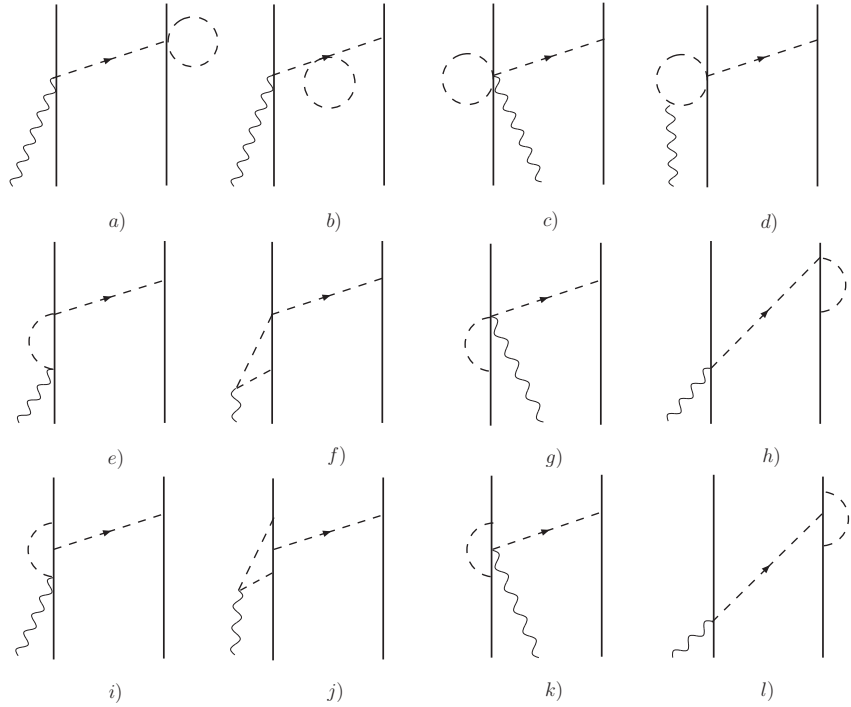}
\caption{Diagrams illustrating loop corrections to tree-level two-body currents, notation
as in Fig.~\protect\ref{fig:fig2}.
Only one among the possible time orderings is shown for each contribution.}
\label{fig:fig5}
\end{figure}
\begin{eqnarray}
{\rm type\,\, a)} &=&
{\bf j}^{\rm NLO}_{\rm b}\,\left[-\frac{3}{2\, F_\pi^2} \, M^{(1)} \right] \ , \\
{\rm type\,\, b)} &=& -i\, e\frac{g^2_A}{F^2_\pi}\,
 ({\bm \tau}_1 \times {\bm \tau}_2)_z \,
 {\bm \sigma}_1 \, \frac{{\bm \sigma}_2\cdot {\bf k}_2}{\omega^2_{k_2}}
\left[-\frac{1}{\omega^2_{k_2}}\,
\frac{m_\pi^2}{F_\pi^2} \,  M^{(1)}\right] \,+ 1 \rightleftharpoons 2 \ , \\
{\rm type\,\, c)} &=& {\bf j}^{\rm NLO}_{\rm b} \,
\left[-\frac{5}{2\, F_\pi^2}\, M^{(1)}\, \right] \ , \\
{\rm type\,\, d)}& =& -i\, e\frac{g^2_A}{2\,F^4_\pi}
 ({\bm \tau}_1 \times {\bm \tau}_2)_z
 \,I^{(2)}_{ij}(q)\,\sigma_{1,j}\, \frac{{\bm \sigma}_2\cdot
{\bf k}_2}{\omega^2_{k_2}}+  1 \rightleftharpoons 2 \ , \\
 {\rm type\,\, e)}& =& {\bf j}^{\rm NLO}_{\rm b} \,
\left[ \frac{1}{F_\pi^2} \,M^{(1)}\, \right] \ ,\\
{\rm type\,\, f)}& =& i\, e\frac{g^2_A}{2\,F^4_\pi}
 ({\bm \tau}_1 \times {\bm \tau}_2)_z
 \,I^{(2)}_{ij}(q)\,\sigma_{1,j}\, \frac{{\bm \sigma}_2\cdot {\bf k}_2}{\omega^2_{k_2}}+  1 \rightleftharpoons 2 \ , \\ 
\label{eq:eqjj} 
{\rm type\,\, j)} &=& i\, e\frac{2\,g^4_A}{F^4_\pi}\, \tau_{2,z}\, J^{(2)}_{ij}(q)\,
({\bf k}_2\times {\bf q})_j\,
\, \frac{{\bm \sigma}_2\cdot {\bf k}_2}{\omega^2_{k_2}} + 1 \rightleftharpoons 2 \ ,\\
 {\rm type\,\,k)} = {\rm type\,\,l)}
&=&{\bf j}^{\rm NLO}_{\rm b} \,
 \left[ \frac{g_A^2}{6\, F_\pi^2} \,M^{(3)}\, \right] \ ,
\label{eq:typer}
\end{eqnarray}
while for those in Fig.~\ref{fig:fig5n}:
\begin{eqnarray}
 {\rm type\,\, m)+type\,\, n)} &=&{\bf j}^{\rm NLO}_{\rm c} \,
\left[-\frac{3}{F_\pi^2} \,M^{(1)}\, \right]\ , \\
 {\rm type\,\, o)+type\,\, p)} &=&{\bf j}^{\rm NLO}_{\rm c}\,
 \left(-\frac{1}{\omega^2_{k_1}}-\frac{1}{\omega^2_{k_2}} \right)
 \frac{m_\pi^2}{F_\pi^2} \, M^{(1)} \, , \\
 {\rm type\,\, q)} &=& {\bf j}^{\rm NLO}_{\rm c} \,
 \left[-\frac{5}{F_\pi^2}\, M^{(1)}\, \right] \ , \\
  {\rm type\,\, r)}& =&  i\, e\frac{g^2_A}{F^4_\pi}\,
 ({\bm \tau}_1 \times {\bm \tau}_2)_z\,I^{(2)}_{ij}(q) \,
({\bf k}_1- {\bf k}_2)_j
\frac{{\bm \sigma}_1\cdot {\bf k}_1}{\omega^2_{k_1}}
\frac{{\bm \sigma}_2\cdot {\bf k}_2}{\omega^2_{k_2}} \ ,\\ 
 {\rm type\,\,u)}+{\rm type\,\, v)}&=& {\bf j}^{\rm NLO}_{\rm c} \,
 \left[ \frac{g_A^2}{3\, F_\pi^2} \,M^{(3)}\, \right] \ ,
\end{eqnarray}
where ${\bf j}^{\rm NLO}_{\rm b}$ and ${\bf j}^{\rm NLO}_{\rm c}$ are
the seagull and pion-in-flight currents of Eqs.~(\ref{eq:nlo1}) and~(\ref{eq:nlo2}),
and the constants $M^{(n)}$, and kernels $I^{(2)}_{ij}(q)$ and $J^{(2)}_{ij}(q)$
are given in Appendix~\ref{app:dimensional}.  The contributions associated
with diagrams of type h), i), s), and t) vanish, since the integrand
is an odd function of the loop momentum ${\bf p}$.  Lastly, diagrams
of type g) are of order $e\, Q^2$~\cite{Pastore08}, and therefore
beyond the order under consideration in the present study ($e\, Q$),
and only a subset of the irreducible diagrams is retained in the
evaluation of the type j) contribution, see Appendix~\ref{app:recoil}.
\begin{figure}[bthp]
\includegraphics[width=4.5in]{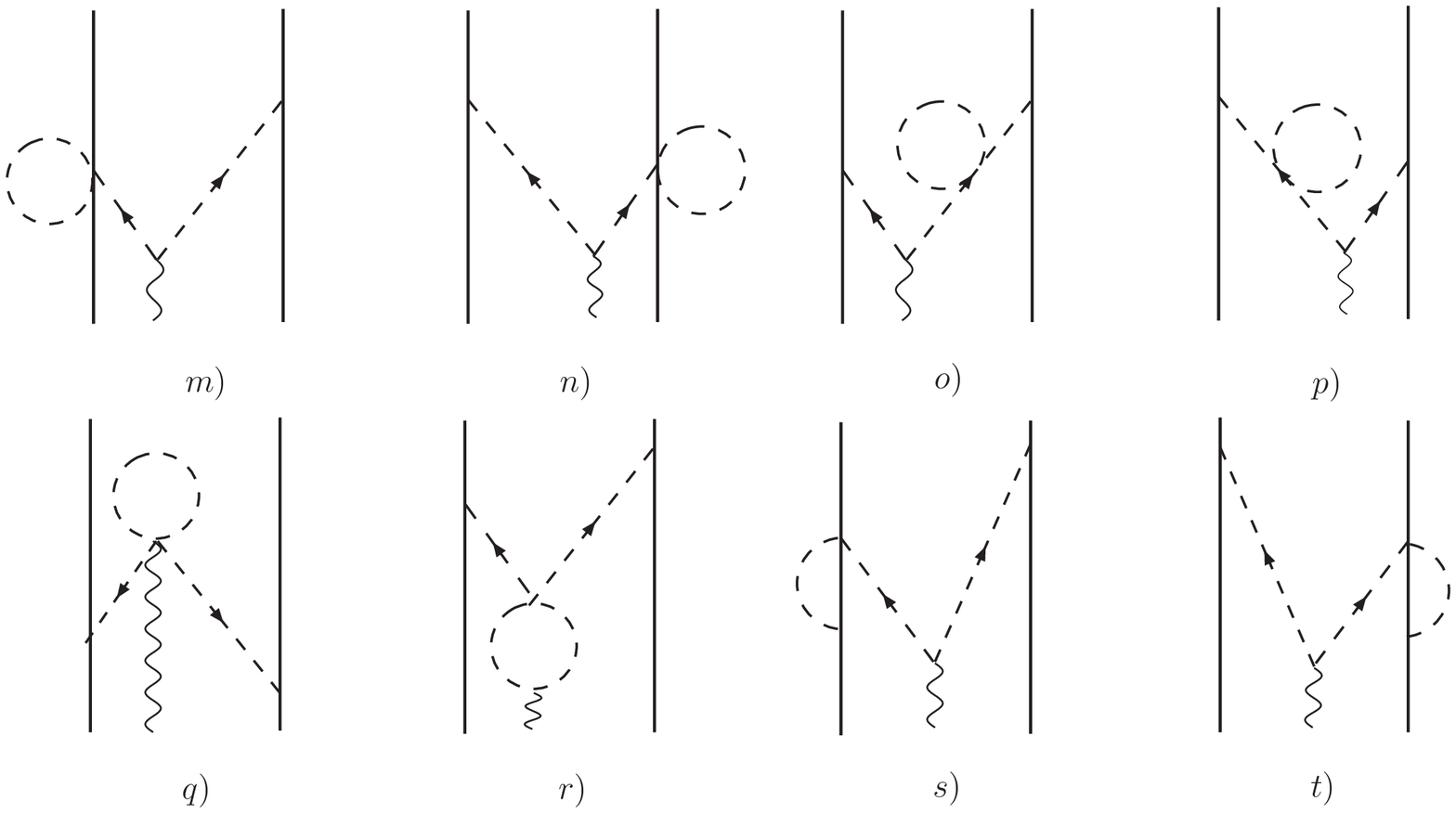}
\includegraphics[width=4.5in]{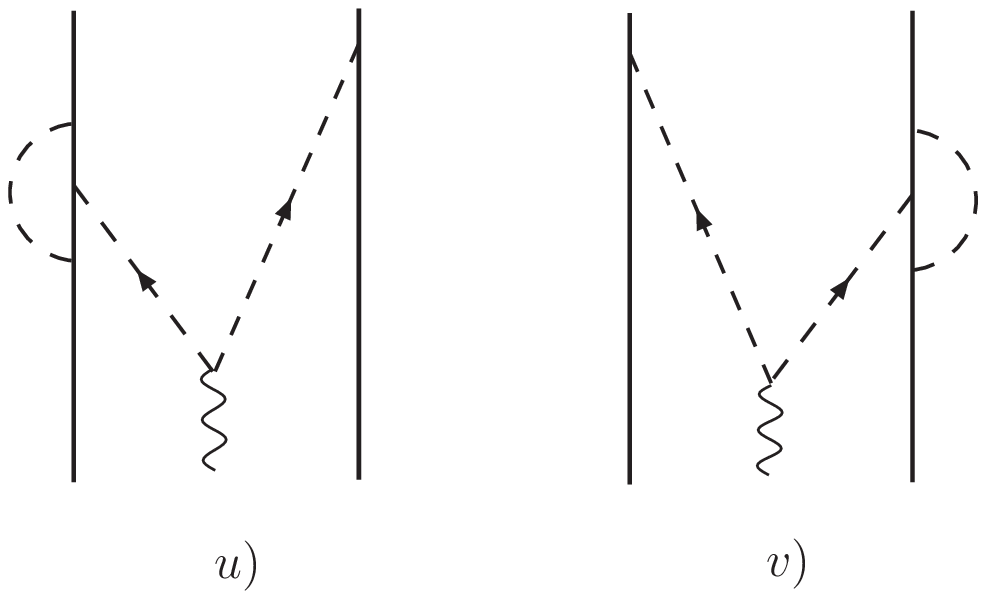}
\caption{Same as in Fig.~\protect\ref{fig:fig5}.}
\label{fig:fig5n}
\end{figure}

A few comments are now in order.  Firstly, the evaluation of the diagrams
in the last row of Figs.~\ref{fig:fig5} and~\ref{fig:fig5n} is carried out
by including recoil corrections to the reducible diagrams of corresponding
topology.  Cancellations occur between the irreducible and these recoil-corrected
reducible contributions.  This is discussed in Appendix~\ref{app:recoil}.

\begin{figure}[bthp]
\includegraphics[width=1.75in]{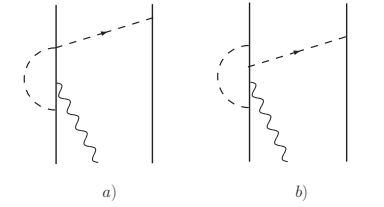}
\caption{Diagrams illustrating N$^4$LO ($e\, Q^{\, 2}$) loop corrections to tree-level currents not
included in the present study, notation as in Fig.~\protect\ref{fig:fig2}.}
\label{fig:fig6}
\end{figure}

Secondly, diagrams like those shown in Fig.~\ref{fig:fig6} have not been
considered since they are, like diagram g) in Fig.~\ref{fig:fig5}, of order $e\, Q^2$,
as can be easily surmised by using the counting rules given in Ref.~\cite{Pastore08}.

Thirdly, the contributions of type a), c), e), k)-l), m)-n), and u)-v) lead to
(further) renormalization of $g_A$, while those of type b) and o)-p) renormalize
the pion mass, namely $\overline{m}^2_\pi=m^2_\pi( 1+ M^{(1)}/F_\pi^2)$.
Thus, both types are accounted for in the (renormalized) seagull and pion-in-flight
currents.  Diagrams j) and r) generate form-factor corrections---their finite
parts follow from the $I^{(2)}_{ij}$ and $J^{(2)}_{ij}$ kernels---to the
nucleon and pion electromagnetic couplings.  However, the contributions
of diagrams d) and f) exactly cancel out.
\section{Magnetic moments from pion exchanges at N$^3$LO}
\label{sec:mum}
To begin with, it is worthwhile making some general considerations.
The magnetic moment operator ${\bm \mu}$ due to a two-body current density ${\bf J}({\bf x})$
can be separated into a term dependent on the center-of-mass position ${\bf R}$
of the two particles and one independent of it~\cite{Dalitz54}, as
\begin{equation}
 {\bm \mu}({\bf R},{\bf r})= \frac{1}{2}\Bigg[ {\bf R} \times \int\!{\rm d}{\bf x}\,\, {\bf J}({\bf x})
+ \int\!{\rm d}{\bf x}\, ({\bf x}-{\bf R}) \times{\bf J}({\bf x}) \Bigg] \ ,
\end{equation}
where, because of translational invariance, ${\bf J}({\bf x})$ is actually a function
of ${\bf J}({\bf x}-{\bf R},{\bf r})$, ${\bf r}$ being the relative position of the
two particles, see Eq.~(\ref{eq:j2rs}).  The first term in square brackets can be related via the continuity
equation to the commutator of the charge density operator with the two-nucleon
potential~\cite{Dalitz54}, assumed to be of the form ${\bm \tau}_1\cdot {\bm \tau}_2
\, V({\bf r})$ but otherwise velocity independent (for example, the one- and two-pion-exchange
potentials derived in Sec.~\ref{sec:potential}), while the second term can be
written in terms of the Fourier transform of ${\bf J}({\bf x})$, denoted by ${\bf j}({\bf q})$.
We find:
\begin{equation}
 {\bm \mu}({\bf R},{\bf r})=-\frac{1}{2}\Big[ e\, ({\bm \tau}_1 \times {\bm \tau}_2)_z\, V({\bf r})\,
{\bf R} \times {\bf r} + i \,\nabla_q \times{\bf j}({\bf q})\!\mid_{{\bf q}=0}\Big]  \ ,
\end{equation}
which, for our purposes, is more conveniently written in momentum space as
\begin{equation}
{\bm \mu}({\bf R},{\bf k})=-\frac{i}{2}\Big[ e\, ({\bm \tau}_1\times {\bm \tau}_2)_z
\,{\bf R}\times \nabla_k \, v({\bf k})+ \nabla_q \times{\bf j}({\bf q},{\bf k})\!
\mid_{{\bf q}=0} \Big] \ ,
\label{eq:mu}
\end{equation}
where $v({\bf k})$ denotes the Fourier transform of $V({\bf r})$.  The first term
above is Sachs' contribution~\cite{Sachs48}, denoted as ${\bm \mu}_{\rm Sachs}$, to
the magnetic moment: it is uniquely determined by the potential between the two nucleons.

Therefore, the currents a)-e) in Fig.~\ref{fig:fig9}
generate a Sachs' magnetic moment---currents g) and i) do not
contribute to it---given by
\begin{equation}
\overline{{\bm \mu}}^{\rm N^3LO}_{\rm Sachs}({\bf R},{\bf k})= -\frac{i}{2}\,
e\, ({\bm \tau}_1\times {\bm \tau}_2)_z
\,{\bf R}\times \nabla_k\, \overline{v}_0^{2\pi}(k) \ ,
\label{eq:musachs}
\end{equation}
where $\overline{v}^{2\pi}_0(k)$
is the term proportional to ${\bm \tau}_1\cdot {\bm \tau}_2$ in Eq.~(\ref{eq:tper}), {\it i.e.}
\begin{equation}
\overline{v}_0^{2\pi}(k)=\frac{1}{48 \pi^2\,F_{\pi}^4}
G(k) \left[4 m_{\pi}^2(1+4 g_A^2-5 g_A^4)+k^2(1+10 g_A^2 - 23 g_A^4)-\frac{48\, g_A^4 m^4_\pi}
{4\,  m^2_\pi+k^2}\right] \ .
\end{equation}
The relation~(\ref{eq:musachs}) can easily be verified by direct
evaluation of $({\bf R}/2)\times {\bf j}_{\rm a-e}({\bf q}=0, {\bf k})$.
The currents a)-e) and i) also generate a translationally invariant
contribution, namely the second term in Eq.~(\ref{eq:mu}), which reads (see
Appendix~\ref{app:muloop} for details)
\begin{equation}
\overline{{\bm \mu}}^{\rm N^3LO}({\bf k}) =\frac{e\,g_A^2}{8\,\pi^2F_{\pi}^4}\,\tau_{2,z}
\Bigg[F_0(k) \,{\bm \sigma}_1 - F_2(k)\, \frac{{\bf k}\,{\bm \sigma}_1\cdot{\bf k}}{k^2}\Bigg]
+ \frac{e\,g_A^2}{2\,\pi^2F_{\pi}^2}\,\tau_{2,z}\,
\left(C_S\, {\bm \sigma}_2-C_T\,{\bm \sigma}_1\right) + 1 \rightleftharpoons 2 \ ,
\label{eq:mufin}
\end{equation}
where the functions $F_i(k)$ are
\begin{eqnarray}
 F_0(k)\!\!&=&\!\!1 -2\, g_A^2+\frac{  8\,g_A^2\, m_\pi^2 }{k^2+4\, m_\pi^2}
+G(k)\left[ 2-2\, g_A^2-\frac{  4\,(1+g_A^2)\, m_\pi^2 }{k^2+4\, m_\pi^2}
+\frac{16\, g_A^2 \,m_\pi^4 }{(k^2+4\, m_\pi^2)^2} \right] \ ,
\label{eq:f0k} \\
 F_2(k)\!\!&=&\!\!2-6\, g_A^2+ \frac{  8\,g_A^2\, m_\pi^2 }{k^2+4\, m_\pi^2}
+G(k)\left[4\, g_A^2-\frac{  4\,(1+3\, g_A^2)\, m_\pi^2 }{k^2+4\, m_\pi^2}
+\frac{16\, g_A^2 \,m_\pi^4 }{(k^2+4\, m_\pi^2)^2} \right] \ . 
\label{eq:f2k}
\end{eqnarray}
It is interesting to note that the constant $2-6\, g_A^2$ in $F_2(k)$
would lead to a long-range contribution of the type
$\left[ \tau_{2,z}\, ({\bm \sigma}_1\cdot {\bm \nabla}){\bm \nabla}
+1 \rightleftharpoons 2 \right]1/r $ in the
magnetic moment, which is, however, fictitious in the
present context of an effective field theory valid at low momenta---in
performing the Fourier transform, the high momentum components
are suppressed by the cutoff $C_\Lambda(k)$.

We now compare the magnetic moment operator derived here with that
obtained in Ref.~\cite{Park96}.   Firstly, we note that the Sachs term
is ignored in that work.  Of course, it vanishes in two-body systems because of
its dependence on ${\bf R}$.  However, in $A>2$ systems the center-of-mass
position of a nucleon pair will not generally coincide with that of the nucleus, and
therefore this term will contribute.

Secondly, the treatment of the box diagrams, panels d) and e) in Fig.~\ref{fig:fig9},
is different in our approach, since the expressions listed in Eqs.~(\ref{eq:mud}) and~(\ref{eq:muef})
result from combining recoil-corrected reducible {\it and} irreducible diagrams.
In particular, had we retained {\it only} the latter, the isospin structure of
${\bm \mu}^{\rm N^3LO}_{\rm d}({\bf k})+{\bm \mu}^{\rm N^3LO}_{\rm e}({\bf k})$
would have contained, in addition to terms proportional to $\tau_{i,z}$, also
the term proportional to $({\bm \tau}_1\times {\bm \tau}_2)_z$ present in Eq.~(46)
of Ref.~\cite{Park96}.

Lastly, we find that type a) and b) contributions in Fig.~\ref{fig:fig9},
which only consist of irreducible diagrams, are in agreement with the corresponding
terms in Eq.~(46) of Ref.~\cite{Park96}.  This is easily seen by considering the
Fourier transform of that equation.  To this end, we first observe that
\begin{equation}
 \int_0^1{\rm d}z\, {\rm ln}\left[1+z(z-1)\, p^2/m_\pi^2 \right] = G(p)-2 \ ,
\end{equation}
and then note that
\begin{eqnarray}
 \int{\rm d}{\bf r}\, {\rm e}^{-i{\bf k}\cdot {\bf r}}\left[ \, r \frac{{\rm d}}{{\rm d}r}
\int_{\bf p} {\rm e}^{i{\bf p}\cdot {\bf r}}\,\left[ 2 -G(p) \right] \right]& =&3\, G(k)+k\, G^{\, \prime}(k)-6 \ , \\
 \int{\rm d}{\bf r}\, {\rm e}^{-i{\bf k}\cdot {\bf r}}\left[ \hat{\bf r}\, 
{\bm \sigma} \cdot \hat{\bf r}\, r\,  \frac{{\rm d}}{{\rm d}r}
\int_{\bf p} {\rm e}^{i{\bf p}\cdot {\bf r}}\,\left[2 -G(p) \right] \right]&=&
\left[G(k)-2\right]\,{\bm \sigma} +k\,G^{\, \prime}(k)\frac{{\bf k}\, 
{\bm \sigma} \cdot {\bf k}}{k^2} \ ,
\end{eqnarray}
where $G^{\,\prime}(k)$ denotes the derivative of $G(k)$.  Inserting these relations
into Eq.~(46) leads to a similar Eq.~(\ref{eq:mufin}), but
with $C_S$ and $C_T$ taken to be zero, and
\begin{eqnarray}
F_0(k)&\rightarrow& G(k)\,\left(2- \frac{4\, m_\pi^2}{k^2+4\, m_\pi^2}\right)-2 \ , \nonumber \\
F_2(k)&\rightarrow& 2-G(k)\,\frac{4\, m_\pi^2}{k^2+4\, m_\pi^2} \nonumber \ .
\end{eqnarray}
The $F_2(k)$ above is the same as Eq.~(\ref{eq:f2k}) (with $g_A$ set to zero to remove
the box contributions), while $F_0(k)$ differs from Eq.~(\ref{eq:f0k}) by a constant, which
gives rise to a zero-ranged operator---operators of this type were dropped in Eq.~(46) anyway.

To the magnetic moment operators of Eqs.~(\ref{eq:musachs}) and~(\ref{eq:mufin}),
one has to add the term of one-pion range originating from the
current ${\bf j}_{\rm tree}^{\rm N^3LO}$ (Sec.~\ref{sec:currents}), given by
\begin{equation}
\overline{\bm \mu}^{\,{\rm N^3LO}}_{\rm tree}=
e\frac{g_A}{F_\pi^2}
\Bigg[
\left(d_8^{\, \prime}\, \tau_{2,z} + 
d_9^{\, \prime}\, {\bm \tau}_1 \cdot {\bm \tau}_2 \right)\,{\bf k}
-d_{21}^{\, \prime} \,({\bm \tau}_1\times{\bm \tau}_2)_z \,{\bm \sigma}_1\times {\bf k}
\Bigg]\,\frac{{\bm \sigma}_2\cdot {\bf k}}{k^2+m_\pi^2}
+ 1 \rightleftharpoons 2 \ .
\end{equation}

\section{Determining the LEC's: fitting the N$^2$LO $N$$N$ potential}
\label{sec:fitp}

We find it convenient to formulate the $N$$N$ scattering-
and bound-state problems in momentum space~\cite{Schiavilla04}.
In the case of scattering, we solve for the $K$-matrix
\begin{equation}
K^{JTS}_{L^\prime,L}(p^\prime,p)=
v^{JTS}_{L^\prime,L}(p^\prime,p)
+\frac{4\mu_N}{\pi} \int_0^\infty {\rm d}kk^2 \sum_{L^{\prime\prime}}
v^{JTS}_{L^\prime,L^{\prime\prime}}(p^\prime,k)
\frac{\cal P}{p^2-k^2}K^{JTS}_{L^{\prime\prime},L}(k,p) \>\>,
\label{eq:kma}
\end{equation}
where $\mu_N$ is the reduced mass, ${\cal P}$ denotes a
principal-value integration, and the momentum-space matrix
elements $v^{JTS}_{L^\prime,L}(p^\prime,p)$ of the potential
are defined as in Eqs.~(3.3) and~(3.4) of Ref.~\cite{Epelbaum00},
but for the factor of $2\pi$ in front of the integration over $z=\hat{\bf p}^\prime
\cdot \hat{\bf p}$ being replaced here by $1/(8\pi)$, and the inclusion, in the present
case, of an additional phase factor $i^{L-L^\prime}$, which, for
coupled channels, leads to mixing angles with signs conforming to the
customary choice made in phase-shift analyses.

The integral equations above are discretized, and the resulting
systems of linear equations are then solved by direct numerical
inversion.  The principal-value integration is removed by
a standard subtraction technique~\cite{Gloeckle83}.  Once the
$K$-matrices in the various channels have been determined, the
corresponding (on-shell) $S$-matrices are obtained from
\begin{equation}
S^{JTS}(p)=\left[ 1+2\, i \, \mu_N p\, K^{JTS}(p,p) \right]^{-1}
\left[ 1-2\, i \, \mu_N p\, K^{JTS}(p,p) \right] \ ,
\label{eq:skma}
\end{equation}
from which phase shifts and, for coupled channels, mixing angles
are easily determined~\cite{Epelbaum00}.

The bound state (with $JTS=101$ and $L,L^\prime=0,2$) is
obtained from solutions of the homogeneous integral equations~\cite{Schiavilla04}
\begin{equation}
w_L(p)=\frac{1}{E_d-p^2/(2\mu_N)} \frac{2}{\pi}
\int_0^\infty {\rm d}k\, k^2 \sum_{L^\prime}
v^{101}_{L,L^\prime}(p,k)\, w_{L^\prime}(k) \ ,
\end{equation}
and from these, for later reference, the configuration-space $S$-
and $D$-wave components follow as
\begin{equation}
u_{L}(r)=\frac{2}{\pi} \int_0^\infty {\rm d}p\, p^2\, j_L(pr)\, w_{L}(p) \ .
\end{equation}

Before turning our attention to a discussion of the phase-shift fits, we note
that the potential constructed in Sec.~\ref{sec:pot_ren} needs to be (further) regularized
because of its power-law behavior for large values of the momenta $k$ and/or
$K$.  This is accomplished by including a high-momentum cutoff, which
we take to be of the form
\begin{equation}
 C_\Lambda(k,K)={\rm e}^{-(k^4+16\,K^4)/\Lambda^4} \ ,
\end{equation}
so that the matrix elements of the regularized potential
entering the $K$-matrix and bound-state equations are
obtained from
\begin{equation}
v^{\rm R}({\bf k},{\bf K})=v({\bf k},{\bf K})\,C_\Lambda(k,K) \ ,
\end{equation}
and $v({\bf k},{\bf K})$ is defined as in Eq.~(\ref{eq:vn2lo}).  In the following
cutoff parameters $\Lambda$ in the range 500--700 MeV are considered.

The LEC's $C_S$, $C_T$, and $C_i$, $i=1,\dots,7$, are determined by fitting
the deuteron binding energy and  S- and P-wave $np$ phase shifts up to
laboratory kinetic energies of 100 MeV, as obtained in the very recent (2008)
analysis of Gross and Stadler~\cite{Gross08}.
\begin{table}
\caption{Values for the nucleon axial coupling constant $g_A$, pion
decay constant $F_\pi$,  neutral and charged pion masses $m_0$ and
$m_+$, (twice) $np$ reduced mass $\mu_N$, and $\hbar c$, used in the fits.}
\begin{tabular}{c|c|c|c|c|c}
\hline
\hline
        $g_A$  & $F_\pi$ (MeV)  & $m_0$ (MeV)  &  $m_+$ (MeV)  & $2\, \mu_N$ (MeV) & $\hbar c$ (MeV-fm)  \\
\colrule
   1.29    &    184.8    &  134.9766      &  139.5702      & 938.9181  & 197.32696    \\
\hline
\end{tabular}
\label{tb:cons}
\end{table}
The parameters characterizing the one- and
two-pion exchange parts of the potential are listed in Table~\ref{tb:cons}, with
the nucleon axial coupling constant $g_A$ determined from the Golberger-Treiman
relation $g_A=g_{\pi NN}F_\pi/(2\, m_N)$, where the $\pi NN$ coupling constant
is taken to have the value $g^2_{\pi NN}/(4\pi)=13.63 \pm 0.20$~\cite{Stoks93a,Arndt94}.
In fact, in the OPE we include the isospin-symmetry breaking induced by the
mass difference between charged and neutral pions, since it leads to significant
effects in the $^1$S$_0$ scattering length~\cite{Wiringa95}, and therefore the OPE potential reads
\begin{equation}
\overline{v}^{\pi}({\bf k})=-\frac{g_A^2}{3\, F_{\pi}^2} \left[ {\bm \tau}_1\cdot{\bm \tau}_2
\left( \frac{1}{k^2+m_0^2}+\frac{2}{k^2+m_+^2} \right)
+T_{12} \left( \frac{1}{k^2+m_0^2}-\frac{1}{k^2+m_+^2} \right)
 \right] {\bm \sigma}_1\cdot{\bf k}\,{\bm\sigma}_2\cdot{\bf k}\ ,
\end{equation}
where $T_{12}$ is the isotensor operator defined as $T_{12}=3\, \tau_{1,z}\tau_{2,z}
-{\bm \tau}_1 \cdot {\bm \tau}_2$, and $m_0$ and $m_+$ are the neutral and charged
pions masses.  Finally, we note that the pion mass entering in the two-pion-exchange part
is taken as $m_\pi=( m_0+2\, m_+)/3$.

The best-fit values obtained for the LEC's are listed in Table~\ref{tb:lec} for $\Lambda$=500, 600,
and 700 MeV.  The fitting strategy becomes obvious once the partial
wave expansion of the potential is carried out.  In the case of spin-singlet ($S=0$)
channels, the contact components of the (partial-wave expanded) potential with $JT$ and $S=0$ read:
\begin{equation}
v^{JT0}_{J,J}(p^\prime,p;{\rm CT0/2})=\frac{1}{8\pi}\int_{-1}^{1} {\rm d}z\,
P_J(z) \left[ D_1+D_2\, (p^{\prime\, 2}+p^2)-2\, D_3\, p^\prime \, p \, z \right]
C_\Lambda(p^\prime,p,z) \ ,
\end{equation}
where $z =\hat{\bf p}^\prime \cdot \hat{\bf p}$, $P_J(z)$ is a Legendre polynomial, and
the $D_i$ denote linear combinations of the LEC's  with
$D_1=C_S-3\, C_T$, $D_2=C_1-3\, C_3-C_6+(C_2-3\, C_4-C_7)/4$, and $D_3=
C_1-3\, C_3-C_6-(C_2-3\, C_4-C_7)/4$.  The cutoff function is even in $z$,
and therefore for even (odd) $J$ only $D_1$ and $D_2$ ($D_3$) contribute.
In practice, $D_1$ and $D_2$ have been determined by fitting the ($n$$p$) singlet
scattering length and effective range, and $^1$S$_0$ phase shift,
while $D_3$ is determined by fitting the $^1$P$_1$ phases.
\begin{table}[bthp]
\caption{Values of the LEC's corresponding to cutoff parameters $\Lambda$ in the range 500--700 MeV,
obtained from fits to $n$$p$ phase shifts up to lab energies of 100 MeV.} \begin{tabular}{c|d|d|d}
\hline
\hline
& \multicolumn{3}{c}{$\Lambda$ (MeV)} \\
\hline
& \multicolumn{1}{c}{500} & \multicolumn{1}{c}{600}  & \multicolumn{1}{c}{700} \\
\hline
$C_S$ (fm$^2$)  &   -4.456420   &   -4.357712   &   -3.863625   \\
$C_T$ (fm$^2$)  &    0.034780   &    0.094149   &    0.234176   \\
$C_1$ (fm$^4$)  &   -0.360939   &   -0.259186   &   -0.268296   \\
$C_2$ (fm$^4$)  &   -1.460509   &   -0.934505   &   -0.835226   \\
$C_3$ (fm$^4$)  &   -0.349780   &   -0.359547   &   -0.389047   \\
$C_4$ (fm$^4$)  &   -1.968636   &   -1.717178   &   -1.724544   \\
$C_5$ (fm$^4$)  &   -0.870067   &   -0.754021   &   -0.695564   \\
$C_6$ (fm$^4$)  &    0.326169   &    0.301194   &    0.348152   \\
$C_7$ (fm$^4$)  &   -0.727797   &   -1.006459   &   -0.955273   \\
\hline
\hline
\end{tabular}
\label{tb:lec}
\end{table}
In the case of spin-triplet ($S=1$) channels, the situation is slightly more complicated.  For uncoupled
channels with $J>0$, we write
\begin{eqnarray}
v^{JT1}_{J,J}(p^\prime,p;{\rm CT0/2})\!\!&=&\!\!\frac{1}{8\pi}\int_{-1}^{1} {\rm d}z\,\Bigg[
P_J(z) \Big[ D_4+(D_5+D_6)\, (p^{\prime\, 2}+p^2)
-2\, (D_7-D_8-D_9)\, p^\prime \, p \, z \Big] \nonumber \\
&-&\Big[ P_{J-1}(z)+P_{J+1}(z)\Big] \,\left(2\,D_8+D_9\right)\, p^\prime \, p \Bigg] \,
C_\Lambda(p^\prime,p,z) \ ,
\end{eqnarray}
while for the $^3$P$_0$ channel (having $JTS=011$)
\begin{eqnarray}
v^{011}_{1,1}(p^\prime,p;{\rm CT0/2})\!\!&=&\!\!\frac{1}{8\pi}\int_{-1}^{1} {\rm d}z\,
\Bigg[ P_1(z) \Big[ D_4+(D_5-D_6)\, (p^{\prime\, 2}+p^2)
-(2\, D_7-D_9)\, p^\prime \, p \, z \Big] \nonumber \\
&+&P_0(z) \,\left(2\, D_8-D_9\right)\, p^\prime \, p \Bigg] \,
C_\Lambda(p^\prime,p,z) \ .
\end{eqnarray}
Here, the $D_i$'s denote the following LEC combinations: $D_4=C_S+C_T$, $D_5= C_1+C_3+(C_2+C_4)/4$,
$D_6=C_6+C_7/4$,  $D_7=C_1+C_3-(C_2+C_4)/4$, $D_8= C_6-C_7/4$, and $D_9=C_5$.  In terms of these, the contact components
for coupled channels are given by
\begin{eqnarray}
v^{JT1}_{- -}(p^\prime,p;{\rm CT0/2})\!\!&=&\!\!\frac{1}{8\pi}
\int_{-1}^{1} {\rm d}z\,
\Bigg[ P_{J-1}(z) \Big[ D_4\!+\!\left(D_5+\frac{D_6}{2J+1}\right) (p^{\prime\, 2}+p^2)
\!-\!\left(2\, D_7-D_9\right) p^\prime \, p \, z \Big] \nonumber \\
&-&P_J(z) \,\left(\frac{2\, D_8}{2J+1}+D_9 \right)\, p^\prime \, p \Bigg] \,
C_\Lambda(p^\prime,p,z) \ ,
\end{eqnarray}
\begin{eqnarray}
v^{JT1}_{++}(p^\prime,p;{\rm CT0/2})\!\!&=&\!\!\frac{1}{8\pi}
\int_{-1}^{1} {\rm d}z\,
\Bigg[ P_{J+1}(z) \Big[ D_4\!+\!\left(D_5-\frac{D_6}{2J+1}\right) (p^{\prime\, 2}+p^2)
\!-\!\left(2\, D_7-D_9\right) p^\prime \, p \, z \Big] \nonumber \\
&+&P_J(z) \,\left( \frac{2\, D_8}{2J+1}-D_9\right)\, p^\prime \, p \Bigg] \,
C_\Lambda(p^\prime,p,z) \ ,
\end{eqnarray}
\begin{eqnarray}
v^{JT1}_{+-}(p^\prime,p;{\rm CT0/2})\!\!&=&\!\!-\frac{1}{4\pi}\frac{\sqrt{J(J+1)}}{2J+1}
\int_{-1}^{1} {\rm d}z\,
\Bigg[ D_6\, \Big[P_{J-1}(z)\, p^{\prime\, 2}+P_{J+1}(z)\, p^2\Big] \nonumber \\
&-& 2\, D_8\, P_J(z)  \, p^\prime \, p \Bigg] \,
C_\Lambda(p^\prime,p,z) \ ,
\end{eqnarray}
where $L=\pm$ is a shorthand for $L=J\pm1$, and the off-diagonal matrix
element with $-+$ is obtained from $v^{JT1}_{+-}(p^\prime,p;{\rm CT0/2})$ by
exchanging $p^\prime \rightleftharpoons p$.  The parameters $D_4$, $D_5$ and $D_6$
are then determined by fitting the deuteron binding energy, spin-triplet scattering length
and effective range, and $^3$S$_1$-$^3$D$_1$ phases and mixing angle---the contributions
of terms proportional to $D_7$, $D_8$, and $D_9$ vanish in this channel.  On the other hand,
only the latter enter into the $^3$P$_{J=0,1,2}$ channels, and the
associated phases can then be used to fit $D_7$, $D_8$, and $D_9$.

\begin{figure}[bthp]
\includegraphics[width=5in]{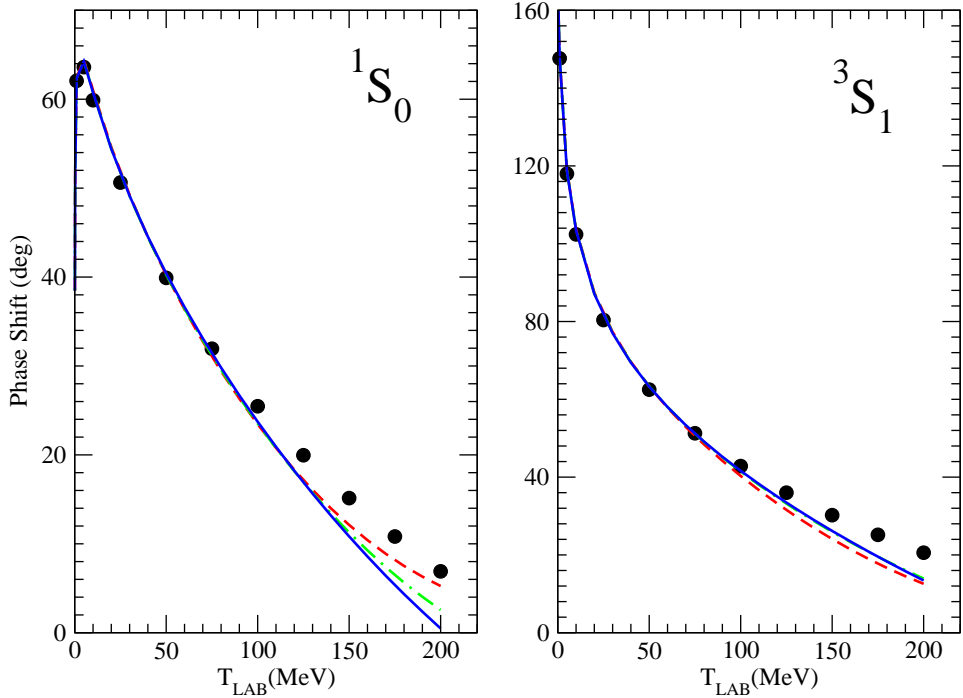}
\caption{(Color online) The S-wave $np$ phase shifts, obtained with cutoff parameters
$\Lambda$=500, 600, and 700 MeV, are denoted by dash (red),
dot-dash (green), and solid (blue) lines, respectively.
The filled circles represent the phase-shift analysis of Ref.~\protect\cite{Gross08}.}
\label{fig:sw}
\end{figure}

\begin{figure}[bthp]
\includegraphics[width=6in]{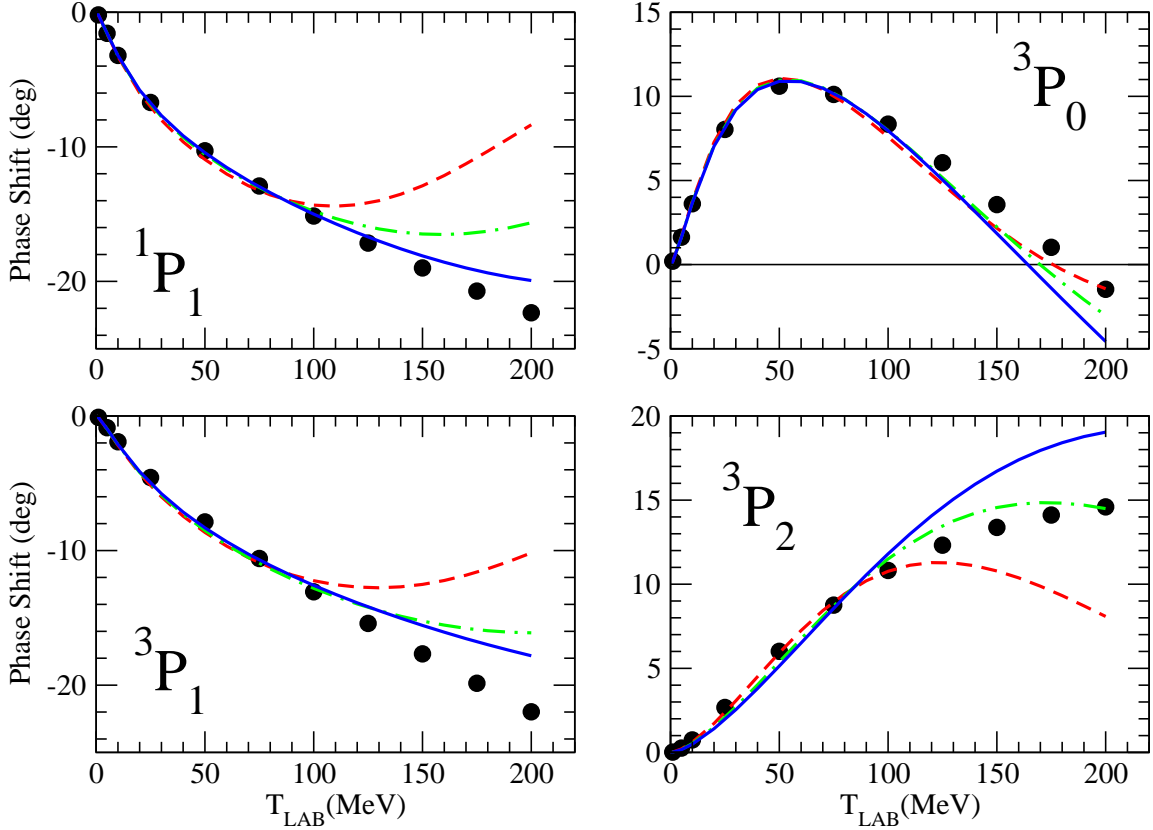}
\caption{(Color online) Same as in Fig.~\protect\ref{fig:sw}, but for P-wave phase shifts.}
\label{fig:pw}
\end{figure}

\begin{figure}[bthp]
\includegraphics[width=6in]{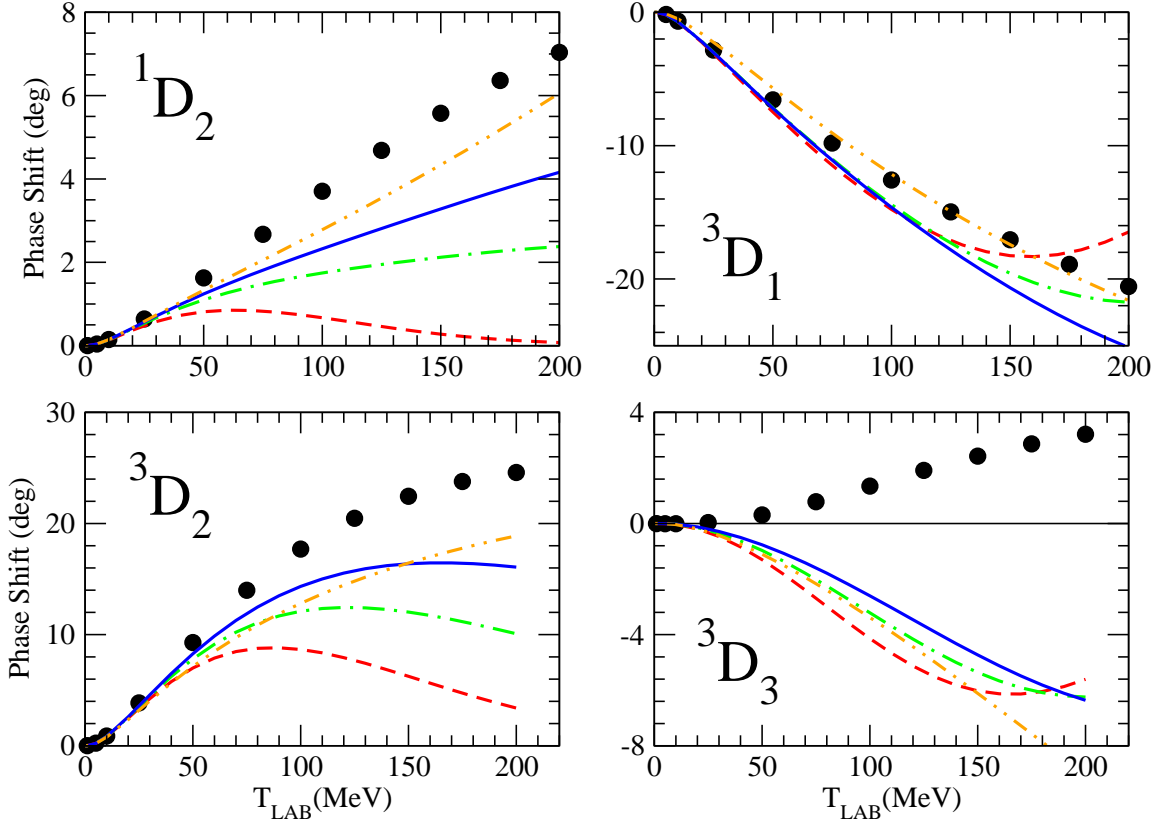}
\caption{(Color online) Same as in Fig.~\protect\ref{fig:sw}, but for D-wave phase shifts.
The dash-double-dot (orange) line
is obtained in first order perturbation theory for the $T$-matrix by including only the
one- and two-pion-exchange parts of the N$^2$LO potential. }
\label{fig:dw}
\end{figure}

\begin{figure}[bthp]
\includegraphics[width=6in]{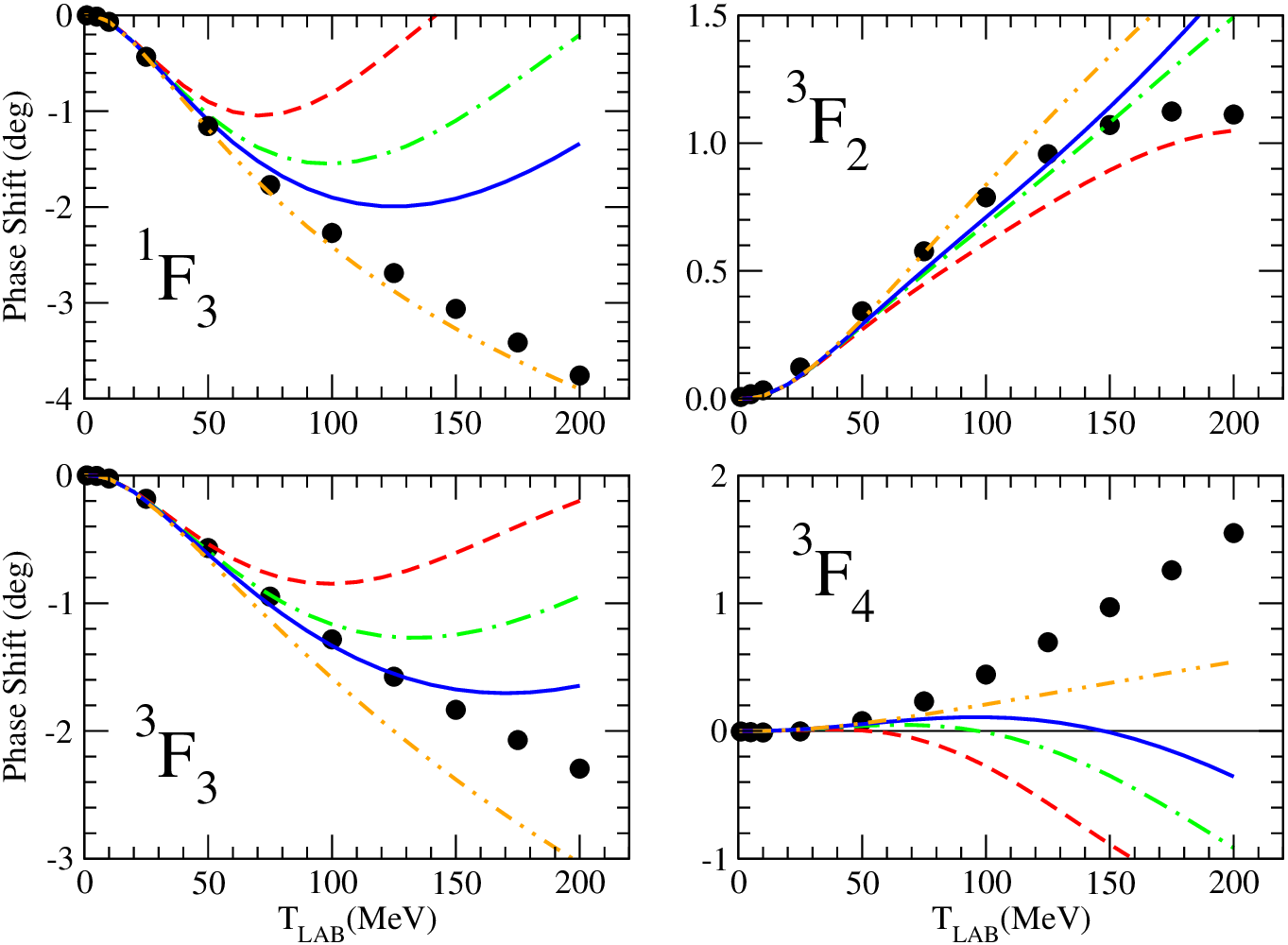}
\caption{(Color online) Same as in Fig.~\protect\ref{fig:dw}, but for F-wave phase shifts. }
\label{fig:fw}
\end{figure}

\begin{figure}[bthp]
\includegraphics[width=6in]{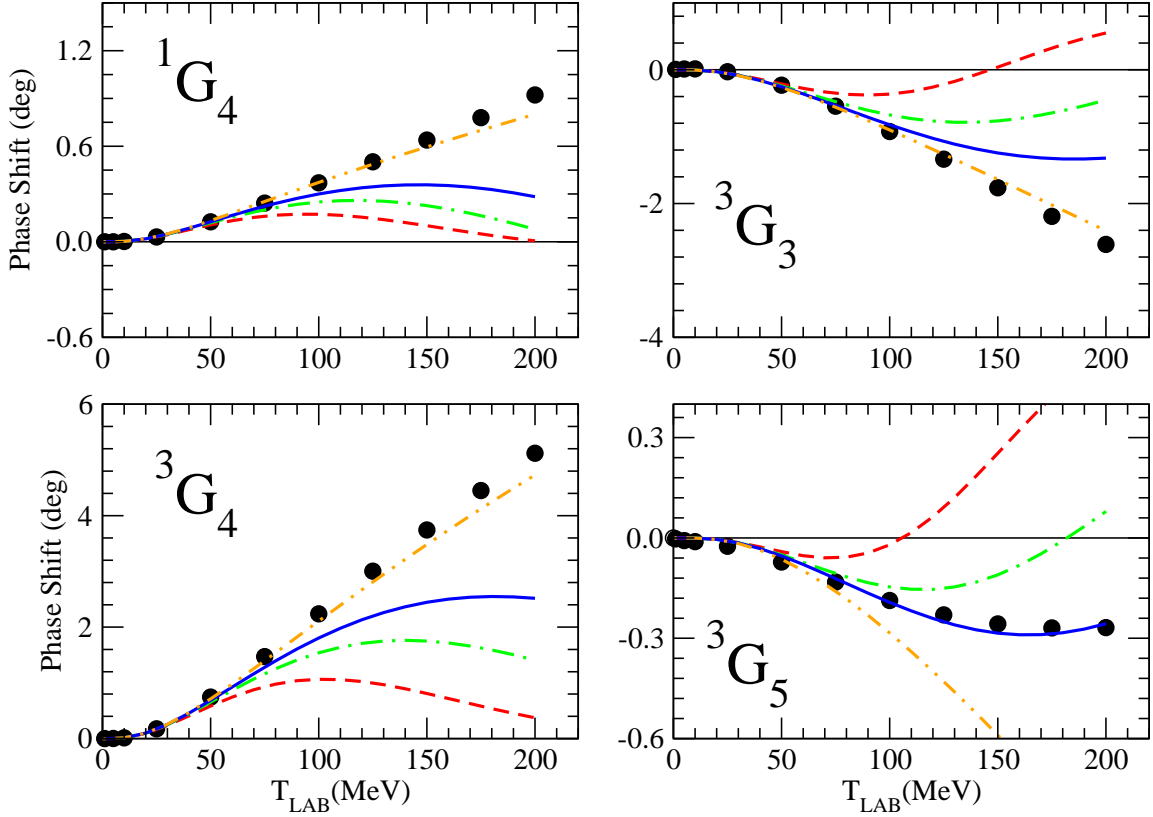}
\caption{(Color online) Same as in Fig.~\protect\ref{fig:dw}, but for G-wave phase shifts.}
\label{fig:gw}
\end{figure}

\begin{figure}[bthp]
\includegraphics[width=6in]{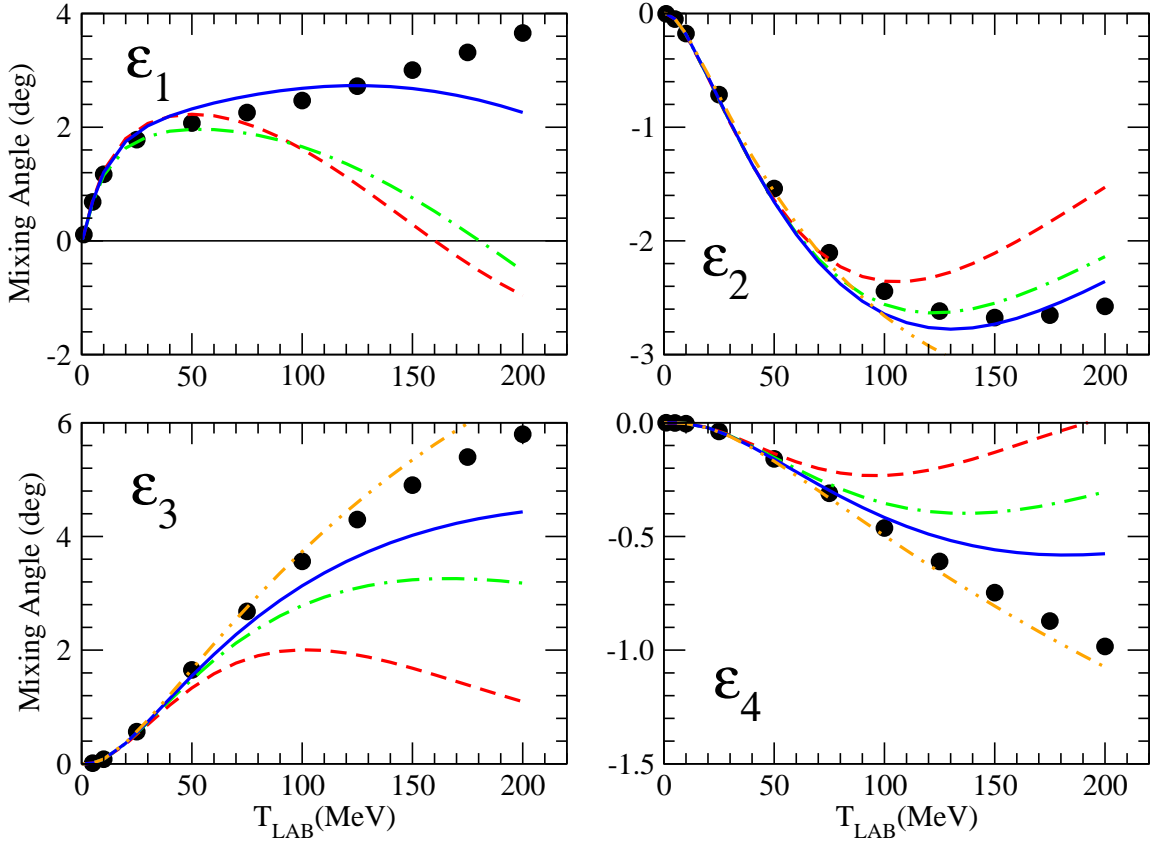}
\caption{(Color online) Same as in Fig.~\protect\ref{fig:dw}, but for the mixing angles $\epsilon_J$.}
\label{fig:ew}
\end{figure}

Results for the S- and P-wave phases used in the fits, as well as for the
D-wave and peripheral  F- and G-wave phases, and mixing angles $\epsilon_{J=1,\dots,4}$
are displayed in Figs.~\ref{fig:sw}--\ref{fig:ew} up to 200 MeV lab kinetic energies.
Effective range expansions and deuteron properties are listed in Table~\ref{tb:erd}.
For reference, in Figs.~\ref{fig:dw}--\ref{fig:ew}, following the
original work by Kaiser {\it et al.}~\cite{Kaiser97}, the phases obtained by including
only the one- and two-pion-exchange ($\overline{v}^\pi$ and
$\overline{v}^{2\pi}$, respectively) terms of the potential are also shown.
These have been calculated in first order perturbation
theory on the $T$-matrix, and hence are cutoff independent.

\vspace{1.25cm}
\begin{figure}[bthp]
\includegraphics[width=6in]{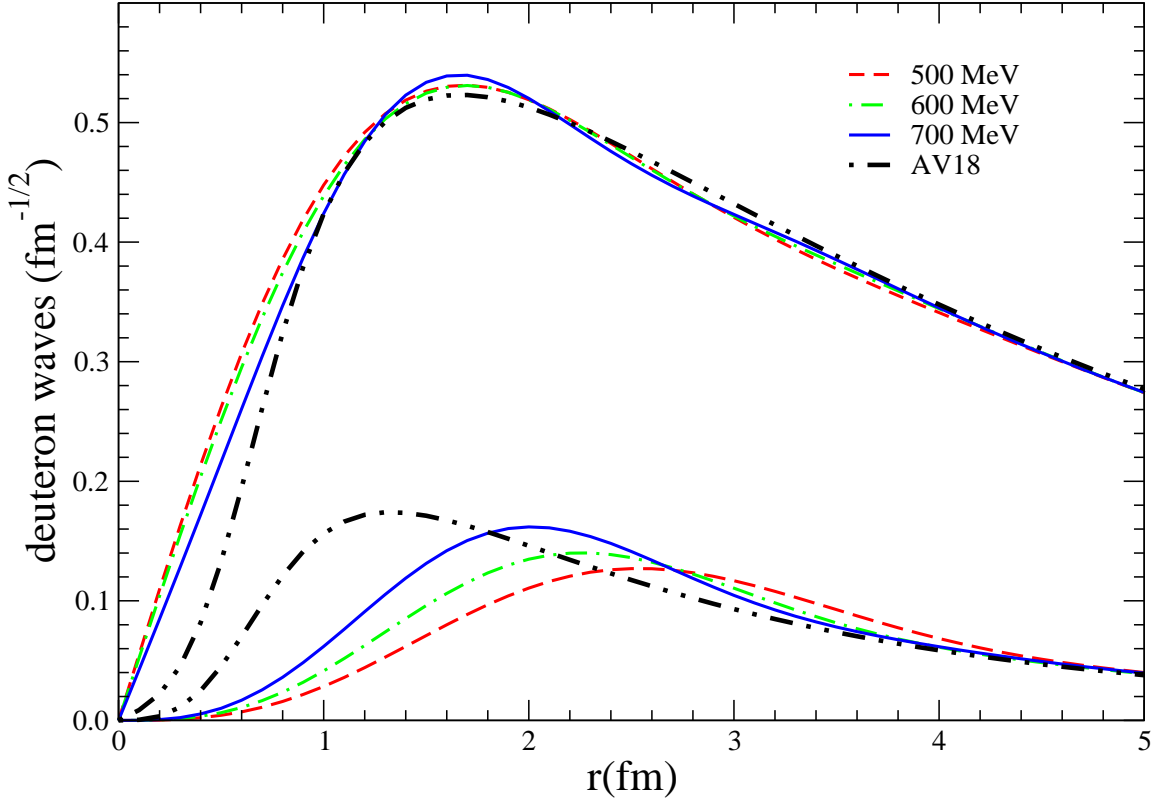}
\caption{(Color online) The S-wave and D-wave components
of the deuteron, obtained with cutoff parameters $\Lambda$=500, 600, and 700 MeV
and denoted by dash (red), dot-dash (green), and solid (blue) lines,
respectively, are compared with those calculated
from the Argonne $v_{18}$ potential (dash-double-dot black lines).}
\label{fig:deut}
\end{figure}
\begin{table}[bthp]
\caption{Singlet and triplet $np$ scattering lengths ($a_s$ and $a_t$) and effective
ranges ($r_s$ and $r_t$), and deuteron binding energy ($B_d$), D- to S-state ratio
($\eta_d$), root-mean-square matter radius ($r_d$), magnetic moment ($\mu_d$), quadrupole
moment ($Q_d$), and D-state probability ($P_D$), obtained
with $\Lambda$=500, 600, and 700 MeV, are compared to the corresponding experimental
values ($a_s$, $r_s$, $a_t$, and $r_t$ from
Ref.~\protect\cite{nper75}, $B_d$ from Ref.~\protect\cite{bd82}, $\eta_d$ from Ref.~\protect\cite{eta90},
$r_d$ and $\mu_d$ from Ref.~\protect\cite{rdmud05}, $Q_d$
from Ref.~\protect\cite{qd79}).}
\begin{tabular}{c|d|d|d|d}
\hline
\hline
& \multicolumn{3}{c}{$\Lambda$ (MeV)} \\
\hline
& \multicolumn{1}{c}{500} & \multicolumn{1}{c}{600}  & \multicolumn{1}{c}{700}
& \multicolumn{1}{c}{Expt} \\
\hline
$a_s$ (fm)         &   -23.729   &   -23.736   &   -23.736   &   -23.749(8)   \\
$r_s$ (fm)         &     2.528   &     2.558   &     2.567   &     2.81(5)    \\
$a_t$ (fm)         &     5.360   &     5.371   &     5.376   &     5.424(3)   \\
$r_t$ (fm)         &     1.665   &     1.680   &     1.687   &     1.760(5)   \\
$B_d$ (MeV)        &     2.2244  &     2.2246  &     2.2245  &     2.224575(9)\\
$\eta_d$           &     0.0267  &     0.0260  &     0.0264  &     0.0256(4)  \\
$r_d$ (fm)         &     1.943   &     1.947   &     1.951   &     1.9734(44) \\
$\mu_d$ ($\mu_N$)  &     0.860   &     0.858   &     0.853   &     0.8574382329(92)\\
$Q_d$ (fm$^2$)     &     0.275   &     0.272   &     0.279   &     0.2859(3)  \\
$P_D$ (\%)         &     3.44    &     3.87    &     4.77    &                \\
\hline
\hline
\end{tabular}
\label{tb:erd}
\end{table}
Overall, the quality of the fits at N$^2$LO
is comparable to that reported in Refs.~\cite{Epelbaum00,Entem02} and,
more recently, in Ref.~\cite{Yang09}.
While the cutoff dependence is relatively weak for the S-wave phases
beyond lab energies of 100 MeV, it becomes significant for higher
partial wave phases and for the mixing angles.  In particular, the F- and G-wave phases,
while small because of the centrifugal barrier, nevertheless display a
pronounced sensitivity to short-range physics, although there are indications~\cite{Kaiser98}
that inclusion of explicit $\Delta$-isobar degrees of freedom might
reduce this sensitivity.
Beyond 100 MeV, the agreement between the calculated and experimental phases
is generally poor, and indeed in the $^3$D$_3$ and $^3$F$_4$ channels
they have opposite sign.  The scattering lengths are well reproduced
by the fits (within $\sim 1$\% of the data, see Table~\ref{tb:erd}),
however, the singlet and triplet effective ranges are both significantly
underpredicted, by $\sim 10$\% and $\sim 5$\% respectively.

The deuteron S- and D-wave radial wave functions are shown in Fig.~\ref{fig:deut}
along with those calculated with the Argonne $v_{18}$ (AV18) potential~\cite{Wiringa95}.
The D wave is particularly sensitive to variations in the
cutoff: it is pushed in as $\Lambda$ is increased from 500 to 700 MeV,
but remains considerably smaller than that of the AV18 up to internucleon
distances of $\sim 1.5$ fm, perhaps not surprisingly, since this realistic
potential has a strong tensor component at short range.  The static
properties, {\it i.e.} D- to S-state ratio, mean-square-root matter
radius, and magnetic moment---the binding energy is fitted---are
close to the experimental values, and their variation with $\Lambda$
is quite modest.  The quadrupole moment is underpredicted by $\sim 4$\%,
a pathology common, to the best of our knowledge, to all realistic potentials
(including the AV18).

\section{N$^3$LO magnetic moment from contact currents}
\label{sec:muct}

The magnetic moment due to the contact currents originating from
minimal couplings (Sec.~\ref{sec:ctcnt}) can also be separated
into a Sachs term and one independent of the center-of-mass position ${\bf R}$
of the two nucleons.  To this end, we first note  that, because of the gradients
acting on the nucleon fields, the $NN$ contact potential contains,
in addition to the contribution $v^{\rm CT2}({\bf k},{\bf K})$ in
Eq.~(\ref{eq:ct2}), also a contribution dependent on the pair momentum
${\bf P}={\bf p}_1+{\bf p}_2={\bf p}^\prime_1+{\bf p}^\prime_2$, given by
\begin{eqnarray}
v_{\bf P}^{\rm CT2}({\bf k},{\bf K})&=&
i\,C^*_1\,\frac{{\bm \sigma}_1-{\bm \sigma}_2}{2}\cdot {\bf P}\times{\bf k}
+C^*_2\,({\bm \sigma}_1\cdot{\bf P}\,\,{\bm \sigma}_2\cdot {\bf K}
-{\bm \sigma}_1\cdot{\bf K}\,\,{\bm \sigma}_2\cdot {\bf P}) \nonumber \\
&+&(C^*_3+C^*_4\,{\bm \sigma}_1\cdot{\bm \sigma}_2)\, P^2
+C^*_5\,{\bm \sigma}_1\cdot{\bf P}\,\,{\bm \sigma}_2\cdot {\bf P} \ ,
\label{eq:ct2p}
\end{eqnarray}
where the $C^*_i$'s consist of the following LEC combinations
\begin{eqnarray}
\label{eq:cs1-5}
C^*_1 &=& C_5^{\prime}/2+C_6^{\prime}/2 \ ,  \nonumber \\
C^*_2 &=& 2\,C_7^{\prime} -2\,C_8^{\prime}-C_{10}^{\prime}
+C_{11}^{\prime} \ , \nonumber \\
C^*_3&=&-C_1^{\prime} +C_2^{\prime}/2-C_{3}^{\prime} \ ,  \\
C^*_4 &=&-C_9^{\prime}+C_{12}^{\prime}/2+C_{14}^{\prime} \ ,\nonumber \\
C^*_5 &=& -C_7^{\prime}/2-C_{8}^{\prime}/2+C_{10}^{\prime}/4
+C_{11}^{\prime}/4+C_{13}^{\prime} \ . \nonumber 
\end{eqnarray}
Incidentally, we observe that Eqs.~(\ref{eq:c1-7}) and~(\ref{eq:cs1-5})
provide a one-to-one correspondence between the LEC's and the coefficients
of the $N\! N$ contact potential.

The (conserved) current ${\rm j}^{\rm N^3LO}_{\rm CT\gamma}$ in
Eq.~(\ref{eq:counter}) gives rise to a Sachs magnetic moment
\begin{equation}
{\bm \mu}^{\rm N^3LO,CT}_{\rm Sachs}=
-\frac{i\, e}{2}\left( 1+\frac{\tau_{1,z}+\tau_{2,z}}{2}\right)
{\bf R}\times \Big[{\bf R}\, , \, v_{\bf P}^{\rm CT2}\Big]
-\frac{i\, e}{4}\frac{\tau_{1,z}-\tau_{2,z}}{2}
{\bf R}\times \Big[{\bf r}\, , \, v^{\rm CT2}+
v_{\bf P}^{\rm CT2}\Big] \ ,
\label{eq:muctc}
\end{equation}
where the only term in $v_{\bf P}^{\rm CT2}$
with a non-vanishing commutator with the relative position
${\bf r}$ is that proportional to $C^*_2$.  Equation~(\ref{eq:muctc})
can be easily verified by considering $({\bf R}/2)\times 
{\bf j}^{\rm N^3LO}_{\rm CT\gamma}({\bf q}=0)$.

The $M1$ operator above depends on the unknown $C^*_i$,
which could be determined, for example, by fitting $A$=3 bound
and scattering state properties, or $M1$ transitions
in light nuclei with $A >2$~\cite{Marcucci08}.
Instead, here we will require that they vanish, {\it i.e.}~that the
contact potential is independent of the nucleon pair momentum.
To the best of our knowledge, this approximation has been
adopted, albeit implicitly, in all studies of $A>2$ nuclei based on $\chi$EFT potentials.
In this respect, we observe that relativistic boost corrections~\cite{Carlson93} to
the rest-frame $v^{\rm CT2}({\bf k},{\bf K})$, being proportional to
$\sim v^{\rm CT2}\, (P^2/m_N^2)$, are suppressed by two additional powers
of the low momentum scale $Q$ relative to both $v^{\rm CT2}({\bf k},{\bf K})$
and $v_{\bf P}^{\rm CT2}({\bf k},{\bf K})$.  These corrections arise
from the relativistic
energy-momentum relation, Lorentz contraction, and Thomas precession
of the spins, and are of a different nature than the ${\bf P}$-dependent
terms in $v_{\bf P}^{\rm CT2}({\bf k},{\bf K})$, which result from
the derivative couplings in the four-nucleon contact Hamiltonians.

Under the assumption above ($C^*_i=0$) and after evaluating the commutator
$[ {\bf r}\, , \, v^{\rm CT2}]$, we find the Sachs magnetic moment to be
given in momentum space by
\begin{eqnarray}
 {\bm \mu}^{\rm N^3LO,CT}_{\rm Sachs}({\bf R},{\bf k},{\bf K})&=&
\frac{e}{4}\frac{\tau_{1,z}-\tau_{2,z}}{2}
{\bf R}\times\Big[ 2\, (C_2 +C_4\, {\bm \sigma}_1\cdot {\bm \sigma}_2)\,
{\bf K} -i\, C_5\, \frac{{\bm \sigma}_1+{\bm \sigma}_2}{2}
\times {\bf k} \nonumber \\
&+&C_7\, 
({\bm \sigma}_1\,{\bm \sigma}_2\cdot {\bf K}
+{\bm \sigma}_1\cdot{\bf K}\,\,{\bm \sigma}_2) \Big] \ .
\end{eqnarray}
It is determined by $C_2$, $C_4$, $C_5$, and $C_7$, {\it i.e.}~by
the LEC's of the momentum-dependent terms in $v^{\rm CT2}$
which do not commute with the charge operator.
In configuration space, ${\bf K}$ reduces to the relative momentum
operator, and the pair correlation function $\delta({\bf r})$ is smeared
over a length scale $1/\Lambda$ ($\Lambda$ is the high-momentum
cutoff introduced in Sec.~\ref{sec:fitp}).

The ${\bf R}$-independent contribution due to minimal
couplings follows from the second term in Eq.~(\ref{eq:mu}),
\begin{equation}
{\bm \mu}^{\rm N^3LO,CT}_{\rm m}
=-\frac{e}{2}(C^\prime_4+C^\prime_5)\,
 ({\bm \sigma}_1+{\bm \sigma}_2) \ ,
\label{eq:min}
\end{equation}
where we have used the relation $C^\prime_6=-C^\prime_5$ implied
by $C^*_1=0$, and have dropped a term proportional to
$(\tau_{1,z}+\tau_{2,z})\, ({\bm \sigma}_1+{\bm \sigma}_2)$,
since it vanishes when acting on antisymmetric two-nucleons states.
However, the contribution due to non-minimal couplings,
which only consists of translationally-invariant terms (the corresponding
currents are transverse to ${\bf q}$ and
therefore unconstrained by the continuity equation), is given by
\begin{equation}
{\bm \mu}^{\rm N^3LO,CT}_{\rm nm}=-e\,
C_{15}^\prime\,({\bm \sigma}_1+{\bm \sigma}_2) -e\, 
C_{16}^\prime\, (\tau_{1,z} -\tau_{2,z}) \,({\bm \sigma}_1-{\bm \sigma}_2) \ .
\label{eq:munm}
\end{equation}
Hence, the $M1$ operator due to minimal and non-minimal couplings is
determined by two independent LEC's, one of which multiplies an isoscalar
structure, while the other multiplies an isovector structure.  The former
(latter) could be determined by reproducing the deuteron magnetic moment
(the cross section for $np$ radiative capture or the isovector
combination of the trinucleon magnetic moment).
\section*{Acknowledgments}
Conversations at various stages of the present work with J.L.\ Goity are
gratefully acknowledged, as is a useful comment by J.D.\ Walecka
in reference to the N$^3$LO magnetic moment operator.  We wish
to thank F.\ Gross and A.\ Stadler for advice relating to their phase-shift
analysis, E.\ Epelbaum for correspondence on various aspects
of the N$^2$LO potential, R.\ Machleidt for a clarification
on a phase convention, and D.R.\ Phillips for a critical
reading of the manuscript.
One of the authors (R.S.) would also like to thank the Physics Department of
the University of Pisa, the INFN Pisa branch, and especially the Pisa group
for the support and warm hospitality extended to him on several occasions.
The work of R.S.\ and R.B.W.\ is supported by the U.S.~Department of Energy,
Office of Nuclear Physics, under contracts DE-AC05-06OR23177 and DE-AC02-06CH11357,
respectively.  Some of  the calculations were made possible by grants of computing
time from the National Energy Research Supercomputer Center.
%
%

\appendix
\section{N$^3$LO currents from non-minimal couplings}
\label{app:nmcounterterms}

External currents enter into the chiral Lagrangian either by the gauging
of spacetime derivatives (minimal coupling), or  through their field
strengths $F_{\mu\nu}$, which transform covariantly under chiral symmetry.
In the case of the electromagnetic current, we have both isoscalar and
isovector components.
In the non-relativistic limit the allowed spin-space structures, at
leading order, are
\begin{equation}
\epsilon_{ijk} F_{ij}\,  
 N^\dagger \sigma_k N\,\, N^\dagger N\ ,
\end{equation}
which, by time-reversal symmetry, can only be associated with the flavor
structures ${\bf 1} \otimes {\bf 1}$, 
$\tau_a \otimes \tau_a$ and ($\tau_z\otimes {\bf 1} \pm 
{\bf 1}\otimes \tau_z$), and
\begin{equation}
 F_{ij}\,  N^\dagger \sigma_i N \,\, N^\dagger  \sigma_j N\ ,
\end{equation}
which can only be associated with the antisymmetric flavor structure
$\epsilon_{zab}\, \tau_a \otimes \tau_b$.
Using the Fierz-type identities for the Pauli matrices,
\begin{eqnarray}
({\bf 1})[{\bf 1}]&=&\frac{1}{2} ({\bf 1}] [{\bf 1}) + \frac{1}{2} 
({\bm \sigma} ]\cdot [{\bm \sigma})\ ,\\
({\bf 1}) [{\bm \sigma}] + ({\bm \sigma})[{\bf 1}]
&=& ({\bf 1}] [{\bm \sigma}) + ({\bm \sigma}][{\bf 1})\ , \\
(\sigma_i)\, [\sigma_j]- (\sigma_j)[\sigma_i]&=&i\,
      \epsilon_{ijk}\Big[(\sigma_k][{\bf 1}) - ({\bf 1}][\sigma_k)\Big]\ ,
\end{eqnarray}
where $(,),[,]$ denote the spinors (or isospinors) $\chi_1^\dagger, \chi_2,
\chi_3^\dagger, \chi_4$, we are left with two operators:
\begin{equation}
H_{\rm CT\gamma,nm}=\frac{e}{2} \int {\rm d}{\bf x}\,
\Bigg[ C_{15}^\prime\,  N^\dagger \sigma_k N \,\, N^\dagger N + 
C_{16}^\prime\,\left( N^\dagger \sigma_k\, \tau_z N \,\, 
N^\dagger N - N^\dagger \sigma_k N \,\,
N^\dagger \tau_z N \right) \Bigg] \epsilon_{ijk}\,F_{ij}\ .
\end{equation}

We also remark that the fourteen operators in the two-nucleon, two-derivative
contact Lagrangian can be reduced to twelve, since, using partial
integration, the following relation involving the vertices
proportional to $C^\prime_4$, $C^\prime_5$ and $C^\prime_6$ and to
$C^\prime_7$, $C^\prime_8$, $C^\prime_{10}$ and $C^\prime_{11}$,  can be shown to hold
\begin{eqnarray}
&&\epsilon_{ijk} \left[ N^\dagger \nabla_i N \,\, (\nabla_j N)^\dagger
\sigma_k N + (\nabla_i N)^\dagger N\,\, N^\dagger \sigma_j \, \nabla_k N
\right] \nonumber \\
&&= \epsilon_{ijk} \left[ N^\dagger \sigma_k N \,\,(\nabla_i
      N)^\dagger \nabla_j N + N^\dagger  N \,\, (\nabla_i
      N)^\dagger \sigma_k \, \nabla_j N \right], \nonumber \\
&&(\delta_{ik} \delta_{jl} - \delta_{il}\delta_{jk}) \left[ N^\dagger \sigma_k\,
  \nabla_i N\,\,  N^\dagger \sigma_l \,\nabla_j N + (\nabla_i N)^\dagger
  \sigma_k N \,\, (\nabla_j N)^\dagger \sigma_l N \right] \nonumber\\
&&= -2 (\delta_{ik} \delta_{jl} - \delta_{il}\delta_{jk})  N^\dagger \sigma_k\,
  \nabla_i N \,\, (\nabla_j N)^\dagger \sigma_l N \ .
\end{eqnarray}
\section{Dimensional Regularization of kernels}
\label{app:dimensional}
In this appendix we report a list of general integration
formulae~\cite{Veltman94,Peskin95},
useful to carry out the regularization of the various kernels occurring in
the potential and current  operators.

\subsection{Useful integrals}

We utilize the Feynman parameterization
\begin{equation}
\label{app:feynmantrick}
 \frac{1}{AB}=\int_0^1 \, {\rm d}y \,\frac{1}{[y A+(1-y)B]^2} \ ,
\end{equation}
and, in order to simplify the energy factors entering the kernels, we make
use of the integral representations~\cite{Rijken92}:
\begin{eqnarray}
& &\frac{1}{\omega_+ + \omega_-}=
\frac{2}{\pi}\int_0^\infty {\rm d}\beta\, \frac{\beta^2}{(\omega_+^2+\beta^2)(\omega_-^2+\beta^2)} \ ,\\
& &\frac{1}{\omega_+\,\omega_-\, (\omega_+ + \omega_-)}=
\frac{2}{\pi}\int_0^\infty {\rm d} \beta\, \frac{1}{(\omega_+^2+\beta^2)(\omega_-^2+\beta^2)} \ .
\end{eqnarray}

Having defined
\begin{equation}
\label{eq:dimreg}
 \int_{\bf p}\, \,\equiv \int \frac{{\rm d}^{d}p}{(2 \pi)^d} \ ,
\end{equation}
we have:
\begin{eqnarray}
\label{app:intl0}
\int_{\bf p} \, \, \frac{1}{(p^2+A)^{\alpha}}
  &=&\frac{1}{(4 \pi)^{d/2}} 
\frac{\Gamma(\alpha-d/2)}{\Gamma(\alpha)}\, A^{-(\alpha-d/2)} \ ,\\
\int_{\bf p}\, \, \frac{p^2}{(p^2+A)^{\alpha}} 
  &=&\frac{1}{(4 \pi)^{d/2}}\,\frac{d}{2}\,\frac{\Gamma(\alpha-d/2-1)}{\Gamma(\alpha)} 
                       \, A^{-(\alpha-d/2-1)} \ ,\\
\int_{\bf p}\, \, \frac{p^4}{(p^2+A)^{\alpha}}   &=&
\frac{1}{(4 \pi)^{d/2}}\,\frac{d\,(d+2)}{4}\,
\frac{\Gamma(\alpha-d/2-2)}{\Gamma(\alpha)} 
                      \, A^{-(\alpha-d/2-2)} \ ,
\end{eqnarray}
where $\Gamma(z)$ is the $\Gamma$-function
satisfying $z\,\Gamma(z)=\Gamma(z+1)$, with
asymptotic behavior for $z\rightarrow 0$ given by
\begin{equation}
 \Gamma(z)=\frac{1}{z} - \gamma +
\left(\frac{\gamma^2}{2}+\frac{\pi^2}{12}\right)z + {\it O}(z^2) \ ,
\end{equation}
and $\gamma\approx0.5772$ is the Euler-Mascheroni constant.  However,
we note that, in order to preserve physical dimensions, a
renormalization scale $\mu$ has to be introduced, and therefore a
factor $\mu^{3-d}$ should be understood in Eq.~(\ref{eq:dimreg}).

Finally, we use the following relations~\cite{Gradshteyn94} to evaluate
\begin{eqnarray}
\label{app:int_table1}
\!\!\!\int\!\! {\rm d}x \ln| x^2-a^2|\!\!&=&\!\! x \ln| x^2-a^2 | - 2\,x + a \ln \left| \frac{x+a}{x-a}\right| \, ,\\
\!\!\!\int\!\! {\rm d}x\,x^2 \ln| x^2-a^2 |\!\!&=&\!\! \frac{1}{3}
\left ( x^3 \ln| x^2-a^2 | - \frac{2}{3}\,x^3 - 2\,a^2 x + a^3 \ln \left| \frac{x+a}{x-a}\right|\right) \, ,\\
\!\!\!\int\!\! {\rm d}x\,x^4 \ln| x^2-a^2 |\!\!&=&\!\! \frac{1}{5}
\left( x^5 \ln| x^2-a^2 | -\frac{2}{5}\,x^5- \frac{2}{3}\,a^2 x^3 - 2 \,a^4 x + a^5 \ln 
\left| \frac{x+a}{x-a}\right|\right) \ .
\end{eqnarray}

\subsection{Regularization of the kernels}
\label{int}

As an example, we sketch the regularization of the kernel $I^{(0)}(k)$, given by
\begin{equation}
 I^{(0)}(k) = \int_{\bf p} \, \,\frac{1}{\omega_+\,\omega_-\, (\omega_+ + \omega_-)}=
\frac{2}{\pi} \int_{\bf p} \,\int_0^\infty {\rm d}\beta\, \frac{1}{(\omega_+^2+\beta^2)(\omega_-^2+\beta^2)} \ ,
\end{equation}
where $\omega_\pm = \sqrt{({\bf p}\pm{\bf k})^2+ 4\, m_{\pi}^2}$.
Using the Feynman integral parameterization of Eq.~(\ref{app:feynmantrick}) with $A=\omega_+^2+\beta^2$ and 
$B = \omega_-^2+\beta^2$, we obtain
\begin{eqnarray}
 I^{(0)}(k) &=&\frac{2}{\pi} \int_{\bf p}\,\int_0^1 {\rm d}y  \,\int_0^\infty {\rm d}\beta\,
\left[ \left[{\bf p}+(2\, y-1)\,{\bf k} \right]^2 +4\, [m_{\pi}^2- y\,(y-1)\,k^2] +\beta^2 \right]^{-2} \nonumber \\
 &=& \frac{1}{2}\, \int_{\bf p} \,\int_0^1 {\rm d} y \left[ p^2+4\,[m_{\pi}^2-y\,(y-1)\,k^2]\right]^{-3/2}\ ,
\end{eqnarray}
where in the second line we have also shifted the integration
variable ${\bf p}\rightarrow{\bf p}+(2\,y -1)\, {\bf k}$.
The integral over ${\bf p}$ is reduced to the form given in Eq.~(\ref{app:intl0})
with $d=3$, $\alpha=3/2$, and $A=4\,\left[ m_{\pi}^2-y\,(y-1)\,k^2\right]$.
With this choice of $d$ and $\alpha$, we are left with a
$\Gamma$-function of vanishing argument.  In order to isolate
the divergent part of the integral, we set $d=3-\epsilon$ and
study its asymptotic behavior for $\epsilon\rightarrow0^+$.  Using
\begin{eqnarray}
\Gamma\left(\frac{\epsilon}{2}\right)&=&\frac{2}{\epsilon}-\gamma +{\it O}(\epsilon) \ ,\\
\Gamma\left(\frac{3}{2}\right)&=&\frac{\sqrt{\pi}}{2} \ ,\\
\left(\frac{A}{4\,\pi}\right)^{-\epsilon/2}&=& 1 -\frac{\epsilon}{2}\ln\frac{A}{4\,\pi} + {\it O}(\epsilon^2)\ ,
\end{eqnarray}
we find, neglecting {\it O}($\epsilon$) terms,
\begin{equation}
\label{app:sigmay}
 I^{(0)}(k)= \frac{1}{8\,\pi^2}\left( \ln
 \pi+\frac{2}{\epsilon}-\gamma \right)
-\frac{1}{8\,\pi^2}\,\int_0^1 {\rm d}y \ln \left[
  \frac{m_{\pi}^2}{\mu^2}-y\,(y-1)\,\frac{k^2}{\mu^2}\right]\ .
\end{equation}
After setting $y \rightarrow (x+1)/2$ and making use of
Eq.~(\ref{app:int_table1}), we obtain:
\begin{equation}
 I^{(0)}(k) = -\,\frac{1}{8\,\pi^2}\left(\frac{s}{k}\ln\frac{s+k}{s-k}
-\frac{2}{\epsilon}+\gamma-\ln{\pi}+\ln \frac{m_{\pi}^2}{\mu^2}-2\right) \ ,
\end{equation}
where $s=\sqrt{4\, m_{\pi}^2+k^2}$.

The kernels
\begin{eqnarray}
 I^{(2)}(k)&=&  \int_{\bf p} \, \,\frac{p^2}{\omega_+\,\omega_-\, (\omega_+ + \omega_-)}\, ,\\
I^{(2)}_{ij}(k)&=& \int_{\bf p} \, \,\frac{p_i\,p_j}{\omega_+\,\omega_-\, (\omega_+ + \omega_-)}\, ,
\end{eqnarray}
can be easily evaluated as shown above.  We find:
\begin{eqnarray}
 I^{(2)}(k)&=&\frac{1}{24\,\pi^2} \Bigg[ \frac{2\,s^3}{k}\ln\frac{s+k}{s-k}+ 2\,k^2 \left(-\frac{2}{\epsilon}+\gamma
-\ln \pi +\ln \frac{m_{\pi}^2}{\mu^2} -\frac{5}{3}\right)  \nonumber \\ 
 &+& 18\,m_{\pi}^2 \left(-\frac{2}{\epsilon}+\gamma-\ln \pi +\ln \frac{m_{\pi}^2}{\mu^2}-\frac{11}{9}\right) \Bigg] \, ,\\
I^{(2)}_{ij}(k)&=&\frac{1}{24\,\pi^2}\, \delta_{ij} \Bigg[ \frac{s^3}{k}\ln\frac{s+k}{s-k}+  k^2\left(-\frac{2}{\epsilon}+\gamma
 -\ln \pi +\ln \frac{m_{\pi}^2}{\mu^2} - 2\right)  \nonumber \\ 
 &+& 6\,m_{\pi}^2 \left(-\frac{2}{\epsilon}+\gamma-\ln \pi +\ln \frac{m_{\pi}^2}{\mu^2}-\frac{5}{3}\right) \Bigg]\,\nonumber\\
&-&\frac{1}{24\,\pi^2}\, \frac{k_i k_j}{k^2} \Bigg[ \frac{s^3}{k}\ln\frac{s+k}{s-k}+k^2\left(-\frac{2}{\epsilon}+\gamma-\ln\pi
+\ln \frac{m_{\pi}^2}{\mu^2} -\frac{8}{3}\right) -8\,m_{\pi}^2\Bigg] \ .
\end{eqnarray}

Next, we note that
\begin{equation}
 f(\omega_+,\omega_-)\equiv \frac{\omega_+^2+\omega_+\,\omega_-+\omega_-^2}
{\omega_+^3\,\omega_-^3(\omega_++\omega_-)}=-\frac{1}{2}\frac{\rm d}{{\rm d}\,m_{\pi}^2}\,
\frac{1}{\omega_+\,\omega_-\, (\omega_+ + \omega_-)}\ ,
\end{equation}
from which we obtain:
\begin{eqnarray}
\!\!\!\!\!\!\!\! J^{(0)}(k) &=&\int_{\bf p} \,f(\omega_+,\omega_-) =
\frac{1}{8\,\pi^2}\, \frac{1}{k\,s}\ln\frac{s+k}{s-k} \ ,\\
\!\!\!\!\!\!\!\!\!\!\!\!J^{(2)}(k) &=&\int_{\bf p} \, p^2 f(\omega_+,\omega_-)
=\!\!-\frac{1}{8\,\pi^2}\left[\frac{2\,s}{k}\ln\frac{s+k}{s-k} +
3 \left(-\frac{2}{\epsilon}+\gamma-\ln\pi+\ln \frac{m_{\pi}^2}{\mu^2}-\frac{2}{3}\right)\right] \ ,\\
\!\!\!\!\!\!\!\!J^{(2)}_{ij}(k)&=&\int_{\bf p} \, p_i p_j f(\omega_+,\omega_-)
=-\frac{1}{8\,\pi^2}\delta_{ij} \left[ \frac{s}{k}\ln\frac{s+k}{s-k}  + \left(-\frac{2}{\epsilon}+\gamma-\ln\pi
+\ln \frac{m_{\pi}^2}{\mu^2}-\frac{4}{3}\right)\right] \nonumber \\
&&\qquad\qquad\qquad\qquad\quad+\frac{1}{8\,\pi^2}\frac{k_i k_j}{k^2}
\left(\frac{s}{k}\ln\frac{s+k}{s-k}-2 \right) \ , 
\label{eq:j2ij} \\
\!\!\!\!\!\!\!\! J^{(4)}(k) &=& \int_{\bf p} \, p^4 f(\omega_+,\omega_-)
=\frac{1}{8\,\pi^2}\Bigg[\frac{8\,s^3}{3\,k}\ln\frac{s+k}{s-k} 
+30 \,m_{\pi}^2 \left(-\frac{2}{\epsilon}+\gamma-\ln\pi +\ln \frac{m_\pi^2}{\mu^2}-\frac{29}{45}\right) \nonumber \\
&&\qquad\qquad\qquad\qquad +\frac{5}{3}\,k^2\left(-\frac{2}{\epsilon}+\gamma-\ln\pi +\ln \frac{m_\pi^2}{\mu^2}-\frac{12}{5}\right) \Bigg] \ .
\end{eqnarray}

The set of kernels involving the energy factor
\[
\frac{2\, \omega_+ + \omega_-}{2\,\omega_+^3\,\omega_-\, (\omega_+ + \omega_-)^2} 
\]
can be reduced to those of type $J^{(2n)}(k)$ by noting that
\begin{equation}
\int_{\bf p}\,\frac{2\, \omega_+ + \omega_-}{2\,\omega_+^3\,\omega_-\, (\omega_+ + \omega_-)^2}
=\frac{1}{4}
\int_{\bf p} \, \frac{\omega_+^2+\omega_+\,\omega_-+\omega_-^2}{\omega_+^3\,\omega_-^3(\omega_++\omega_-)}
=\frac{1}{4}\, J^{(0)}(k) \ ,
\end{equation}
and similarly for $J^{(2)}(k)$, $ J^{(2)}_{ij}(k)$, $ J^{(4)}(k)$.

The kernels involving the energy factor $g(\omega_+,\omega_-)$,
\begin{eqnarray}
g(\omega_+,\omega_-)&=&\frac{3}{2}\,\frac{2\,\omega_+ +\omega_-}{\omega_+^5\,\omega_-(\omega_+ +\omega_-)^2}+
\frac{\omega_++2\, \omega_-}{\omega_+^3\,\omega_-^3(\omega_+ +\omega_-)^2} \nonumber\\
&=&-\frac{1}{2}\,\frac{\rm d}{{\rm d}\,m^2_{\pi}}\frac{2\, \omega_+ + \omega_-}{2\,\omega_+^3\,\omega_-\, (\omega_+ + \omega_-)^2}\ ,
\end{eqnarray}
easily follow from
\begin{equation}
K^{(0)}(k)=\int_{\bf p}\,g(\omega_+,\omega_-) =-\frac{1}{8}\frac{\rm d}{{\rm d }m_{\pi}^2}\,J^{(0)}(k)=
\frac{1}{16}\frac{{\rm d}^2}{{\rm d} (m_{\pi}^2)^2}\,I^{(0)}(k)\ ,
\end{equation}
and similarly for $K^{(2n)}(k)$, leading to:
\begin{eqnarray}
K^{(0)}(k)&=&\int_{\bf p}\,g(\omega_+,\omega_-)=\frac{1}{64\pi^2}\left[\frac{2}{k\,s^3} \ln{\frac{s+k}{s-k}}+\frac{1}{s^2\,m_{\pi}^2}\right] \, ,\\
K^{(2)}(k)&=& \int_{\bf p}\,p^2\, g(\omega_+,\omega_-)=\frac{1}{64\pi^2}
\left[\frac{4}{k\,s} \ln{\frac{s+k}{s-k}}+\frac{1}{m_{\pi}^2}\right] \, ,\\
K^{(2)}_{ij}(k)&=&\int_{\bf p}\, p_{i}\, p_{j} \,g(\omega_+,\omega_-)\nonumber \\
&=& \frac{1}{64\pi^2}\delta_{ij} \left[\frac{2}{k\,s} \ln{\frac{s+k}{s-k}}\right]
-\frac{1}{64\pi^2}\,\frac{k_i k_j}{k^2}\left[ \frac{2}{k\,s} \ln{\frac{s+k}{s-k}}-\frac{1}{m_{\pi}^2}\right]\, ,\\
K^{(4)}(k)&=& \int_{\bf p}\,p^4\, g(\omega_+,\omega_-)  \nonumber\\
&=&-\frac{1}{64\pi^2}
\left[\frac{16\,s}{k} \ln{\frac{s+k}{s-k}}-\frac{k^2}{m_{\pi}^2}
+30
\left(-\frac{2}{\epsilon}+\gamma-\ln{\pi}+\ln \frac{m_{\pi}^2}{\mu^2} \right)\right] \, ,\\
K^{(4)}_{ij}(k)&=&\int_{\bf p}\,p^2\,p_i\, p_j \,g(\omega_+,\omega_-) \nonumber \\
&=&-\frac{1}{64\pi^2}\,\delta_{ij}
\left[\frac{8\,s}{k} \ln{\frac{s+k}{s-k}}+10
\left(-\frac{2}{\epsilon}+\gamma-\ln{\pi}+\ln \frac{m_{\pi}^2}{\mu^2} -\frac{8}{15}\right)\right]  \nonumber\\
&+& \frac{1}{64\pi^2}\frac{k_ik_j}{k^2}\left[\frac{8\,s}{k} \ln{\frac{s+k}{s-k}}+\frac{k^2}{m_{\pi}^2}-16\right]\,.
\end{eqnarray}

Finally, for the kernel entering diagram e) in Fig.~\ref{fig:fig1}, we obtain
\begin{eqnarray}
L(k) &=&\int_{\bf p}\frac{(\omega_+-\omega_-)^2}{\omega_+\,\omega_-(\omega_++\omega_-)}=
\int_{\bf p}\left[-\frac{4}{(\omega_++\omega_-)} + \frac{2}{\omega_+}\right] \nonumber \\
&=&-\frac{1}{6\,{\pi}^2}\Bigg[\frac{s^3}{k}\,\ln \frac{s+k}{s-k}-8\,m_{\pi}^2 
+ k^2\left(-\frac{2}{\epsilon}+\gamma -\ln \pi +\ln \frac{m_{\pi}^2}{\mu^2} -\frac{8}{3}\right)\Bigg] \ ,
\end{eqnarray}
while for the constants $M^{(n)}$ entering Eqs.~(\ref{eq:vgi})--(\ref{eq:vgh}),
\begin{eqnarray}
M^{(1)}&=&\int_{\bf p}\frac{1}{\omega_p}=\frac{m_{\pi}^2}{8\,\pi^2}
\left(-\frac{2}{\epsilon}+\gamma -\ln 4 \pi +\ln \frac{m_{\pi}^2}{\mu^2}-1\right) \ ,\\
M^{(3)}&=&\int_{\bf p}\frac{p^2}
{\omega_p^3}=\frac{3\, m_{\pi}^2}{8\,\pi^2}\left(-\frac{2}{\epsilon}+\gamma -\ln 4 \pi 
+\ln \frac{m_{\pi}^2}{\mu^2}-\frac{1}{3}\right)\ .
\end{eqnarray}

\section{One-loop two-body currents}
\label{app:cntold}

In this appendix we list the expressions for the one-loop currents
derived in Ref.~\cite{Pastore08}.  Referring to Fig.~\ref{fig:fig9}, we
have:
\begin{eqnarray}
{\rm type\,\, a)}&=&-2\,i\frac{e\, g^2_A}{F^4_\pi} \,
 \int \frac{2\, \tau_{2,z}\,({\bm \sigma}_1 \times {\bf q}_2) \, 
 +({\bm \tau}_1\times{\bm \tau}_2)_z\, {\bf q}_2}
{\omega_1 \, \omega_2 (\omega_1 +\omega_2) }+ 1 \rightleftharpoons 2   \ ,
 \label{eq:dia_a} \\
{\rm type\,\, b)}&=& 2\, i\frac{e\,g^2_A}{F^4_\pi} \,
 \int \frac{{\bf q}_1-{\bf q}_3}{\omega_1\, \omega_2\, \omega_3}
\frac{\omega_1+\omega_2+\omega_3}
  {(\omega_1+\omega_2)(\omega_1+\omega_3)(\omega_2+\omega_3)} \Big[({\bm \tau}_1\times{\bm \tau}_2)_z
\,{\bf q}_1\cdot{\bf q}_2 \nonumber\\
&-&2\, \tau_{2,z}\,{\bm \sigma}_1\cdot({\bf q}_1\times{\bf q}_2)\Big]
+ 1 \rightleftharpoons 2 \  ,\\
{\rm type\,\, c)}&=&- i \frac{e}{2\, F^4_\pi} \, 
\left({\bm \tau}_1 \times {\bm \tau}_2\right)_z
 \int \frac{{\bf q}_1-{\bf q}_3}{\omega_1\,\omega_3} 
\frac{\omega_2(\omega_1+\omega_2+\omega_3)-3\, \omega_1\,\omega_3}
{(\omega_1+\omega_2)(\omega_1+\omega_3)(\omega_2+\omega_3)} \ ,
\\
{\rm type\,\, d)}&=&-2\, i\frac{e\,g^4_A}{F^4_\pi}
 \int \frac{\omega_1^2+\omega_2^2 +\omega_1 \omega_2}{\omega_1^3\,\omega_2^3\,(\omega_1+\omega_2)}\,
 \Big[ ({\bm \tau}_1\times{\bm \tau}_2)_z \,{\bf q}_2\, ({\bf q}_1\cdot{\bf q}_2)  
 + 2 \, \tau_{2,z}\, {\bf q}_1\cdot{\bf q}_2 \, ({\bm \sigma}_1\times{\bf q}_2) \nonumber \\
 \!\!& + &\!\! 2 \, \tau_{1,z}\,{\bf q}_2\, {\bm \sigma}_2\cdot ({\bf q}_1\times{\bf q}_2) \Big] + 1 \rightleftharpoons 2  \ ,
 \label{eq:dia_e} \\
{\rm type \,\, e)}&=&2\, i\frac{e\,g^4_A}{F^4_\pi}
 \int \,({\bf q}_1-{\bf q}_3)
  f(\omega_1,\omega_2,\omega_3)\,
 \Big[ ({\bm \tau}_1\times{\bm \tau}_2)_z \, ({\bf q}_1\cdot{\bf q}_2) ({\bf q}_2\cdot{\bf q}_3) \nonumber \\
 &+& 2 \, \tau_{2,z}\, ({\bf q}_2\cdot{\bf q}_3) \, {\bm \sigma}_1\cdot ({\bf q}_2\times{\bf q}_1) 
  +  2 \, \tau_{1,z}\, ({\bf q}_1\cdot{\bf q}_2) \, {\bm \sigma}_2\cdot ({\bf q}_3\times{\bf q}_2) \Big] \ ,
\label{eq:dia_f} \\
{\rm type \,\, g)}&=&2\, i\frac{e\, g^2_A\, C_T}{F^2_\pi}\, 
({\bm \tau}_1\times{\bm \tau}_2)_z\,
 \int \frac{{\bf q}_1-{\bf q}_2}{\omega_1^3\, \omega_2^3} \frac{\omega_1^2+\omega_1\, \omega_2+\omega_2^2}
 {\omega_1+\omega_2}({\bm \sigma}_1
\cdot{\bf q}_2)({\bm \sigma}_2\cdot{\bf q}_1)  \ ,\\
{\rm type \,\, i)}&=&i\frac{e\,g^2_A}{ F^2_\pi}\, 
\tau_{1,z}\,
 \int \frac{{\bf q}_1-{\bf q}_2}{\omega_1^3\, \omega_2^3} \frac{\omega_1^2+\omega_1\, \omega_2+\omega_2^2}
 {\omega_1+\omega_2} \Bigg[ C_{S}\, {\bm \sigma}_1\cdot({\bf q}_1\times{\bf q}_2) \nonumber \\
&-&\, C_{T}\, {\bm \sigma}_2\cdot({\bf q}_1\times{\bf q}_2) \Bigg] 
+ 1 \rightleftharpoons 2 \ ,
\end{eqnarray} 
where the ${\bf q}_i$'s and $\omega_i=(q_i^2+m_\pi^2)^{1/2}$
denote the momenta (with the flow as indicated in the figure) and
energies of the exchanged pions, and the integration is on any one
of the ${\bf q}_i$'s, the remaining ${\bf q}_j$'s with $j\not= i$
being fixed by momentum-conserving $\delta$-functions.
Lastly, the function $f(\omega_1,\omega_2,\omega_3)$ in the
type e) current is defined as
\begin{eqnarray}
\label{eq:effe}
f(\omega_1,\omega_2,\omega_3)&=&  \frac{1}{\omega_1\, \omega_2\, \omega_3
(\omega_1+\omega_2)(\omega_1+\omega_3)(\omega_2+\omega_3)} 
\Bigg[\frac{\omega_1\,\omega_2+\omega_2\,\omega_3+\omega_1\,\omega_3}{\omega_1\,\omega_2\,\omega_3} \nonumber \\
&+&\frac{ (\omega_1+\omega_2)\,(\omega_2+\omega_3)\,(\omega_1^2+\omega_3^2)}
 {\omega_1^2\, \omega_2\, \omega_3^2}
+\frac{\omega_2}{\omega_1\,\omega_3} 
+\frac{\omega_1+\omega_2+\omega_3}{\omega_2^2}\Bigg] \ .
\end{eqnarray}
\section{Magnetic moments from loop currents}
\label{app:muloop}

In this appendix we list the translationally invariant contributions to the magnetic
moment---second term in Eq.~(\ref{eq:mu})---associated with currents a)-e) and i)
in Fig.~\ref{fig:fig9}.  The contributions of currents c) and g) vanish, while those
of currents a), d), and i) read:
\begin{eqnarray}
\overline{{\bm \mu}}^{\rm N^3LO}_{\rm a}({\bf k})
&=&\frac{e\,g_A^2}{8\,\pi^2F_{\pi}^4}\,\tau_{2,z}G(k)
\Bigg[\left(1-\frac{2\,m_{\pi}^2}{4\,m_{\pi}^2+k^2}\right){\bm \sigma}_1
+\frac{2\,m_{\pi}^2}{4\,m_{\pi}^2+k^2}\frac{{\bf k}\,{\bm \sigma}_1\cdot{\bf k}}{k^2}\Bigg]\nonumber \\
&+&\frac{e\,g_A^2}{8\, \pi^2F_{\pi}^4}\,\tau_{2,z}\left( {\bm \sigma}_1-
\frac{{\bf k}\,{\bm \sigma}_1\cdot{\bf k}}{k^2}\right)
+ 1 \rightleftharpoons 2 \ , 
\label{eq:mua}\\
\overline{{\bm \mu}}^{\rm N^3LO}_{\rm d}({\bf k})&=&
-\frac{e\,g_A^4}{8 \pi ^2\,F_{\pi}^4}\tau_{2,z}\,G(k)
\Bigg[\Bigg[1 -\frac{2\, m_{\pi}^2}{4\,m^2_{\pi}+k^2}-
\frac{8\, m_{\pi}^4}{(4\,m^2_{\pi}+k^2)^2}\Bigg]\,{\bm \sigma}_1\nonumber \\
&+&\Bigg[-\frac{2\,m_{\pi}^2}{4\,m^2_{\pi}+k^2}+\frac{8\, m_{\pi}^4}
{(4\,m^2_{\pi}+k^2)^2}\Bigg] \frac{{\bf k}\, {\bm \sigma}_1\cdot{\bf k}}{k^2}\Bigg] 
 -\frac{e\,g_A^4}{8 \pi ^2\,F_{\pi}^4}\tau_{2,z}
\Bigg[ \left(1- \frac{4\,m_\pi^2}{4\, m_\pi^2+k^2}\right) {\bm \sigma}_1 \nonumber \\
&-& \left(1- \frac{4\,m_\pi^2}{4\, m_\pi^2+k^2}\right) \frac{{\bf k}\,
{\bm \sigma}_1\cdot{\bf k}}{k^2}\Bigg]+ 1 \rightleftharpoons 2  \ ,
\label{eq:mud}\\
\overline{{\bm \mu}}^{\rm N^3LO}_{\rm i}({\bf k})&=&
\frac{e\,g_A^2}{2 \pi ^2\,F_{\pi}^2}\, \tau_{1,z}\, 
\left(C_S \,{\bm \sigma}_1 - C_T\,{\bm \sigma}_2 \right)+ 1 \rightleftharpoons 2  \ .
\end{eqnarray}

Finally, in terms of the kernels $J^{(n)}$ and $K^{(n)}$, the contributions resulting
from currents b) and e) are given by
\begin{equation}
 {\bm \mu}^{\rm N^3LO}_{\rm b}({\bf k})=
\frac{e\,g_A^2}{2\,F_{\pi}^4}\,\tau_{2,z}\,
\Bigg[ \left[J^{(2)}_{ij}(k) -k_i\, k_j\,J^{(0)}(k)\right]\sigma_{1,j}
-\left[J^{(2)}(k)-k^2J^{(0)}(k)\right]{\bm \sigma}_1\Bigg]+ 1 \rightleftharpoons 2  \ ,
\label{eq:mub}
\end{equation}
\begin{eqnarray}
{\bm \mu}^{\rm N^3LO}_{\rm e}({\bf k})&=&\frac{2\,e\,g_A^4}{F_{\pi}^4}\,
\tau_{2,z}\Bigg[
\Big[K^{(4)}(k)-2\, k^2 K^{(2)}(k)+k^4 K^{(0)}(k)\Big] {\bm \sigma}_1 
-4\,\epsilon_{i jk}\,k_k \,({\bm \sigma}_1\times {\bf k})_l K^{(2)}_{jl}(k)\nonumber \\
&-&\Big[K^{(4)}_{ij}(k)-k^2 K^{(2)}_{ij}(k)-k_ik_jK^{(2)}(k)+k_ik_j k^2 K^{(0)}(k)\Big]\,\sigma_{1,j}
\Bigg]+ 1 \rightleftharpoons 2 \  ,
\label{eq:mue}
\end{eqnarray}
from which the renormalized operators follow as
\begin{eqnarray}
\overline{{\bm \mu}}^{\rm N^3LO}_{\rm b}({\bf k}) &=&\frac{e\,g_A^2}{8\,\pi^2F_{\pi}^4}\,\tau_{2,z}G(k)
\Bigg[\left(1-\frac{2\,m_{\pi}^2}{4\,m_{\pi}^2+k^2}\right){\bm \sigma}_1
+\frac{2\,m_{\pi}^2}{4\,m_{\pi}^2+k^2}\frac{{\bf k}\,{\bm \sigma}_1\cdot{\bf k}}{k^2}\Bigg]\nonumber \\
&-&\frac{e\,g_A^2}{8\, \pi^2F_{\pi}^4}\,\tau_{2,z}\frac{{\bf k}\,{\bm \sigma}_1\cdot{\bf k}}{k^2}
+ 1 \rightleftharpoons 2 \ , 
\label{eq:mubf}\\
\overline{{\bm \mu}}^{\rm N^3LO}_{\rm e}({\bf k})  &=&-\frac{e\,g_A^4}{8 \pi ^2\,F_{\pi}^4}\tau_{2,z}\,G(k)
\Bigg[\Bigg[1 +\frac{6\, m_{\pi}^2}{4\,m^2_{\pi}+k^2}-
\frac{8\, m_{\pi}^4}{(4\,m^2_{\pi}+k^2)^2}\Bigg]\,{\bm \sigma}_1\nonumber \\
&+&\Bigg[4-\frac{10\,m_{\pi}^2}{4\,m^2_{\pi}+k^2}+\frac{8\, m_{\pi}^4}
{(4\,m^2_{\pi}+k^2)^2}\Bigg] \frac{{\bf k}\, {\bm \sigma}_1\cdot{\bf k}}{k^2}\Bigg] 
 -\frac{e\,g_A^4}{8 \pi ^2\,F_{\pi}^4}\tau_{2,z}
\Bigg[ \left(1- \frac{4\,m_\pi^2}{4\, m_\pi^2+k^2}\right) {\bm \sigma}_1 \nonumber \\
&-& \left(5- \frac{4\,m_\pi^2}{4\, m_\pi^2+k^2}\right) \frac{{\bf k}\,
{\bm \sigma}_1\cdot{\bf k}}{k^2}\Bigg]+ 1 \rightleftharpoons 2  \ .
\label{eq:muef}
\end{eqnarray}
\section{Recoil corrections}
\label{app:recoil}

\begin{figure}[bthp]
\includegraphics[width=5in]{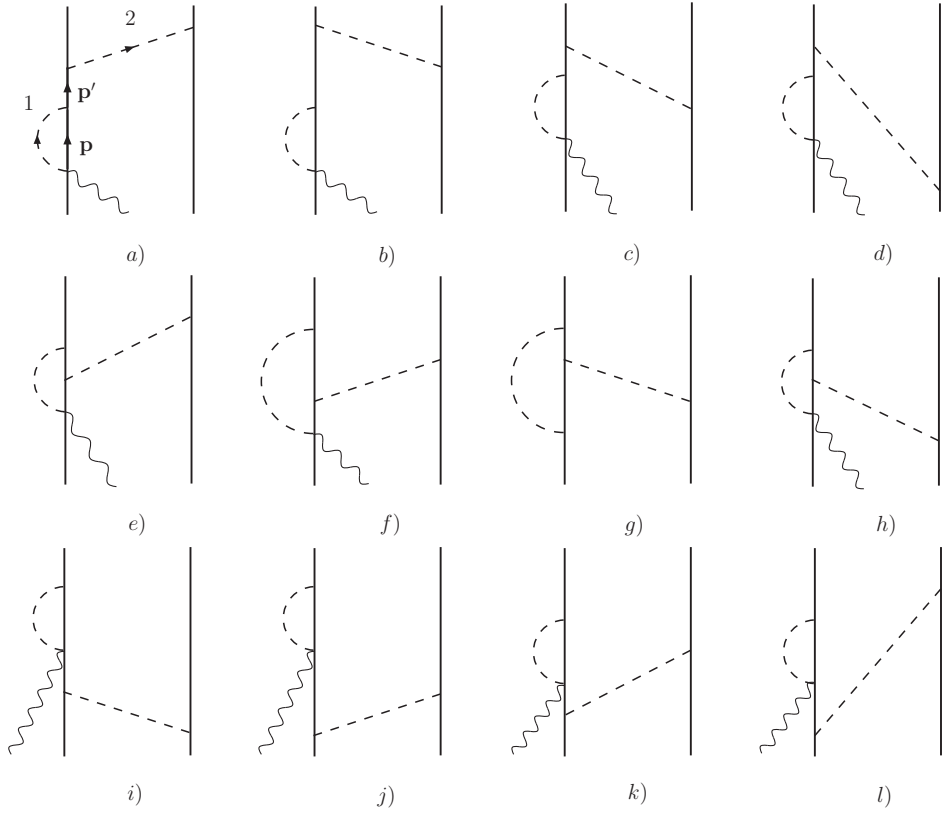}
\caption{Set of time-ordered diagrams for the contribution illustrated by the
single diagram i) in Fig.~\protect\ref{fig:fig5}.  Notation as in Fig.~\protect\ref{fig:fig2}.}
\label{fig:fig7}
\end{figure}

Consider the set of time-ordered diagrams, displayed in
Fig.~\ref{fig:fig7} and denoted as type i) in Fig.~\ref{fig:fig5}.  It is easily seen that recoil corrections in
diagrams a)+b) and i)+j) cancel out the contributions associated
with diagrams c)+d) and k)+l), respectively, so that the expression
for type i) diagrams in Fig.~\ref{fig:fig5}---which happens to vanish---results from
diagrams e)-h).  Let $N$ denote the product
of the four vertices in diagrams a)-d); then the contribution of
diagrams a)+b) is given by
\begin{eqnarray}
 {\rm a)+b)\,\, of\,\, Fig.~14}&=& \frac{N}{(E_i-E_p^\prime -E_2+i\eta)
(E_i-E_p -E_2-\omega_1+i\eta)} \times \nonumber \\
&& \left[ \frac{1}{E_i-E_1^\prime -E_2-\omega_2+i\eta} 
+ \frac{1}{E_i-E_p^\prime -E^\prime_2-\omega_2+i\eta}\right] \ ,
\end{eqnarray}
where the labeling of the momenta is as in panel a), and $E_p$ and $E_p^\prime$
are the energies of the intermediate nucleons.  The expression
in square brackets above can be expanded as
\begin{equation}
 \Bigg[\dots\Bigg] \simeq -\frac{1}{\omega_2} \left[ 2 +
\frac{E_i-E_p^\prime-E_2}{\omega_2} \right] \ ,
\end{equation}
where use has been made of (overall) energy
conservation, $E_i=E_1^\prime+E_2^\prime$, and hence
\begin{equation}
{\rm a)+b)\,\, of \,\, Fig.~14}=
({\rm terms\,\, in\,\, iterated\,\, LS\,\, equation}) -\frac{N}
{\omega_2^2\,(E_i-E_p -E_2-\omega_1+i\eta)} \ .
\end{equation}
\begin{figure}[bthp]
\includegraphics[width=5in]{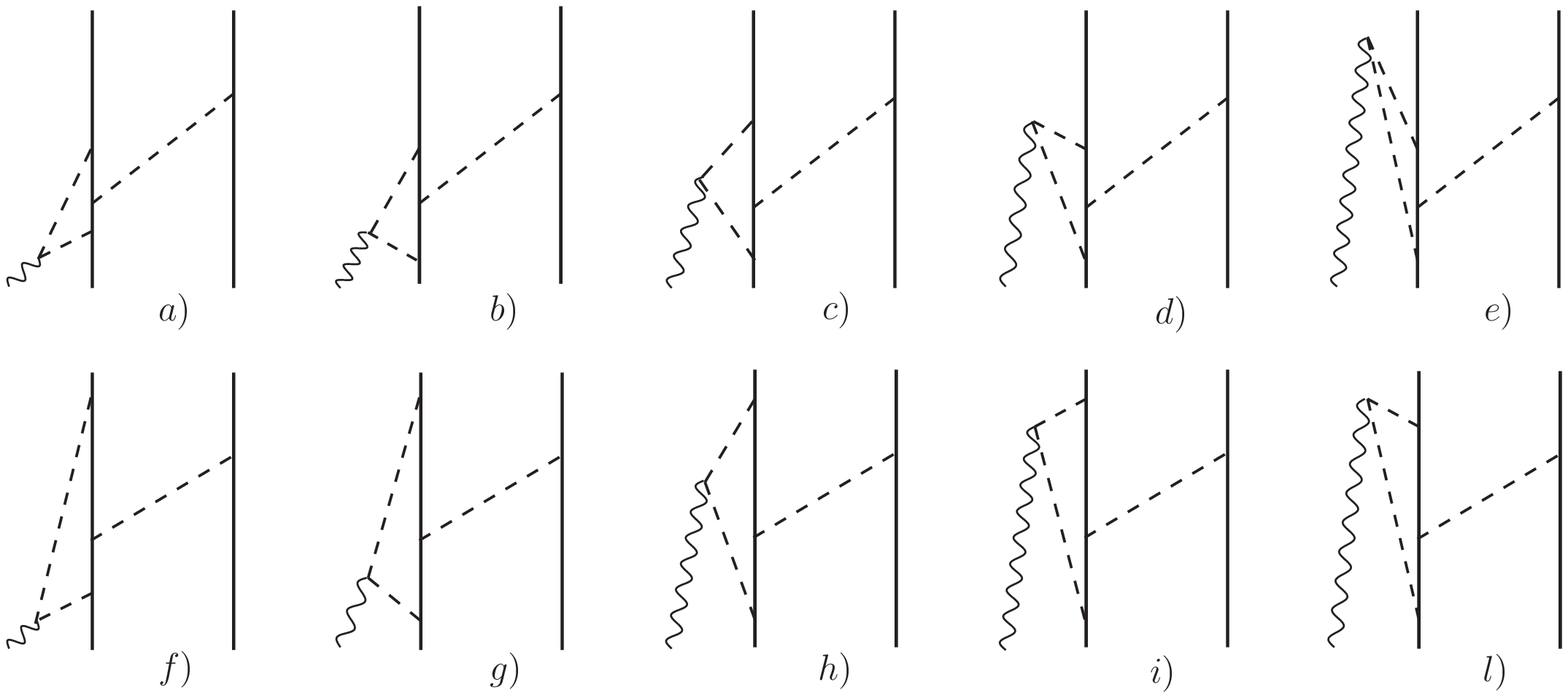}
\includegraphics[width=5in]{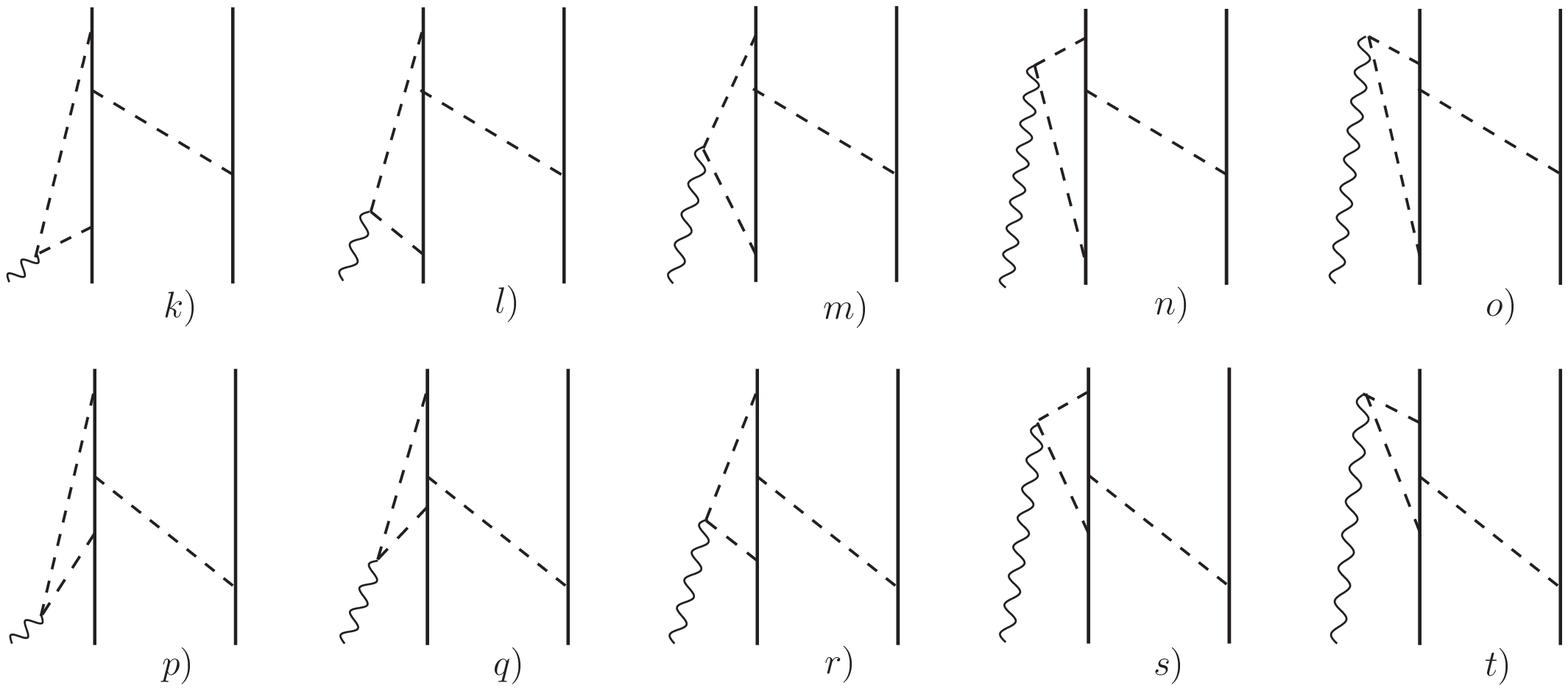}
\caption{Subset of time-ordered diagrams for the contribution illustrated
by the single diagram j) in Fig.~\protect\ref{fig:fig5}.  See text for discussion.
Notation as in Fig.~\protect\ref{fig:fig2}.}
\label{fig:fig8}
\end{figure}
The second term above in the static limit reduces to $N/(\omega_1 \, \omega_2^2)$,
which exactly cancels the contribution of diagrams c)+d).  These exact cancellations
persist also in the k)-l) as well as u)-v) type diagrams of Figs.~\ref{fig:fig5} and~\ref{fig:fig5n},
so that in computing their contributions we only take into account the subset of
(twenty, see below) time-ordered diagrams of topology as shown in those figures.

For the type j) contribution we find that the cancellation between irreducible
and recoil-corrected reducible diagrams is only partial, and the result given
in Eq.~(\ref{eq:eqjj}) corresponds to taking into account only the irreducible
diagrams illustrated in Fig.~\ref{fig:fig8} (the same subset considered in the
evaluation of type u)-v) above).  However, the remaining irreducible and
recoil-corrected reducible diagrams produce an additional contribution of
the form
\begin{eqnarray}
\label{eq:eqjjc} 
&& e\frac{g^4_A}{F^4_\pi}N_{ij}(q) \Bigg[ ({\bm \tau}_1\times{\bm \tau}_2)_z \,
({\bf q}\times {\bf k}_2)_j +\tau_{2,z}\left[{\bf q}\times 
({\bm \sigma}_1\times {\bf k}_2)\right]_j\Bigg]\, \frac{{\bm \sigma}_2\cdot {\bf k}_2}{\omega^2_{k_2}}  \nonumber \\
&+&e\frac{g^4_A}{2\,F^4_\pi}N_{i}(q)\,  \tau_{2,z}\,
\frac{{\bm \sigma}_1\cdot {\bf k}_2 \,\,{\bm \sigma}_2\cdot {\bf k}_2}{\omega^2_{k_2}}
 + 1 \rightleftharpoons 2 \ ,
\end{eqnarray}
where the kernels $N_{ij}$ and $N_i$ are
\begin{eqnarray}
N_{ij}(q)&=&\int_{\bf p} \, \,\frac{p_i\, p_j}{\omega^2_+\,\omega^2_-\, (\omega_+ + \omega_-)} \ , \\
N_{i}(q)&=&\int_{\bf p} \, p_i \,(p^2-q^2) \,\frac{\omega_+-\omega_-}
{\omega^2_+\,\omega^2_-\, (\omega_+ + \omega_-)^2} \ ,
\end{eqnarray}
which, however, does not lead to a Hermitian current density, since this would require
${\bf j}({\bf k}_1,{\bf k}_2)={\bf j}^\dagger(-{\bf k}_1,-{\bf k}_2)$.  We have ignored
this contribution.
\end{document}